\documentclass[11pt]{report} 

\usepackage[utf8]{inputenc} 

\usepackage{geometry} 
\usepackage{graphicx} 

\usepackage{booktabs} 
\usepackage{array} 
\usepackage{paralist} 
\usepackage{verbatim} 
\usepackage{subfig} 

\usepackage{fancyhdr} 
\usepackage{sectsty}
\allsectionsfont{\sffamily\mdseries\upshape} 

\usepackage[nottoc,notlof,notlot]{tocbibind} 
\usepackage[titles,subfigure]{tocloft} 

\usepackage{amsmath,amssymb,amsthm}
\usepackage{gensymb}
\usepackage{cite}
\usepackage[numbers]{natbib}
\usepackage{authblk}
\usepackage{graphicx}
\usepackage{epstopdf}
\usepackage{appendix}

\DeclareMathOperator{\Tr}{Tr}
\DeclareMathOperator{\lb}{\langle\langle}
\DeclareMathOperator{\rb}{\rangle\rangle}
\DeclareMathOperator*{\argmin}{\arg\!\min}

\newcommand*{\Cdot}{\raisebox{-0.25ex}{\scalebox{1.2}{$\cdot$}}}

\def\@fnsymbol#1{\ensuremath{\ifcase#1\or *\or \dagger\or \ddagger\or
   \mathsection\or \mathparagraph\or \|\or **\or \dagger\dagger
   \or \ddagger\ddagger \else\@ctrerr\fi}}

\renewcommand*{\thefootnote}{\fnsymbol{footnote}}

\title{Introduction to Quantum Gate Set Tomography}
\author[1,2]{Daniel Greenbaum\thanks{greenbaum.daniel@gmail.com}}
\affil[1]{Altamira Technologies Corporation, 8201 Greensboro Drive, Suite 800, McLean, VA 22102, USA}
\affil[2]{Laboratory for Physical Sciences, College Park, MD 20740, USA}

\begin{document}
\maketitle
\tableofcontents

\renewcommand*{\thefootnote}{\arabic{footnote}}

\chapter{Introduction}

Progress in qubit technology requires accurate and reliable methods for qubit characterization. There are several reasons characterization is important. One is diagnostics. Qubit operations are susceptible to various types of errors due to imperfect control pulses, qubit-qubit couplings (crosstalk), and environmental noise. In order to improve qubit performance, it is necessary to identify the types and magnitudes of these errors and reduce them. Another reason is the desirability to have metrics of gate quality that are both platform independent and provide sufficient descriptive power to enable the assessment of qubit performance under a variety of conditions. Metrics are also necessary to ascertain if the requirements of quantum error correction (QEC) are being met, i.e., whether the gate error rates are below a suitable QEC threshold. Full characterization of quantum processes provides detailed information on gate errors as well as metrics of qubit quality such as gate fidelity.

Several methods of qubit characterization are currently available. In chronological order of their development, the main techniques are quantum state tomography (QST)~\cite{leibfried_experimental_1996}, quantum process tomography (QPT)~\cite{chuang_prescription_1997,poyatos_complete_1997}, randomized benchmarking (RB)~\cite{knill_randomized_2008,ryan_randomized_2009,magesan_scalable_2011,magesan_characterizing_2012}, and quantum gate set tomography (GST)~\cite{Merkel2013,RBK2013}. All of these tools, with the exception of GST, have been well studied and systematized, and have gained widespread acceptance and use in the quantum computing research community.

GST grew out of QPT, but is somewhat more demanding in terms of the number of experiments required as well as the post-processing. As we will see, obtaining a GST estimate involves solving a highly nonlinear optimization problem. In addition, the scaling with system size is polynomially worse than QPT because of the need to characterize multiple gates at once. Approximately 80 experiments are required for a single qubit and over 4,000 for 2 qubits to estimate a complete gate set, compared to 16 and 256 experiments respectively to reconstruct a single 1- or 2-qubit gate with QPT. Methods for streamlining the resource requirements for GST are under investigation~\cite{IBMpcomm,RBK-pcomm}. Nevertheless, single-qubit GST has been demonstrated by several groups~\cite{Merkel2013,RBK2013} and 2-qubit GST is widely believed to be achievable. Importantly, these groups have convincingly shown that GST outperforms QPT in situations relevant to fault-tolerant quantum information processing (QIP)~\cite{Merkel2013,RBK2013}. As a result, it seems clear that GST in either its present or some future form will supercede QPT as the only accurate method available to fully characterize qubits for fault-tolerant quantum computing. In this document, we present GST in the hopes that readers will be inspired to implement it with their qubits and explore the possibilities it offers.

GST arose from the observation that QPT is inaccurate in the presence of state preparation and measurement (SPAM) errors. 
It will be useful to classify SPAM errors into two different types, which we will call {\em intrinsic} and {\em extrinsic}. Intrinsic SPAM errors are those that are inherent in the state preparation and measurement process. One example is an error initializing the $|0\rangle$ state due to thermal populations of excited states. Another is dark counts when attempting to measure, say, the $|1\rangle$ state. Extrinsic SPAM errors are those due to errors in the gates used to transform the initial state to the starting state (or set of states) for the experiment to be performed. In QPT, the starting states must form an informationally complete basis of the Hilbert-Schmidt space on which the gate being estimated acts. These are typically created by applying gates to a given initial state, usually the $|0\rangle$ state, and these gates themselves may be faulty.

Intrinsic SPAM errors are of particular relevance to fault-tolerant quantum computing, since it turns out that QEC requirements are much more stringent on gates than on SPAM. According to recent results from IBM~\cite{IBMpcomm}, a 50-fold increase in intrinsic SPAM error reduces the surface code threshold by only a factor of 3-4. Therefore QPT - the accuracy of which degrades with increasing SPAM - would not be able to determine if a qubit meets threshold requirements when the ratio of intrinsic SPAM to gate error is large.

This is not an issue for extrinsic SPAM errors, which go to zero as the errors on the gates go to zero. Nevertheless, extrinisic SPAM error interferes with diagnostics: as an example, QPT cannot distinguish an over-rotation error on a single gate from the same error on all gates (see Sec.~\ref{examples-subsection}). In addition, Merkel, et al. have found that, for a broad range of gate error -- including the thresholds of leading QEC code candidates -- the ratio of QPT estimation error to gate error {\em increases} as the gate error itself decreases~\cite{Merkel2013}. This makes QPT less reliable as gate quality improves.

Extrinsic SPAM error is also unsatisfactory from a theoretical point of view: QPT assumes the ability to perfectly prepare a complete set of states and measurements. In reality, these states and measurements are prepared using the same faulty gates that QPT attempts to characterize. One would like to have a characterization technique that takes account of SPAM gates self-consistently. We shall see that GST is able to resolve all of these issues.

Another approach to dealing with SPAM errors is provided by randomized benchmarking~\cite{knill_randomized_2008,ryan_randomized_2009,magesan_scalable_2011,magesan_characterizing_2012}. RB is based on the idea of twirling~\cite{bennett_mixed-state_1996,dankert_exact_2009} -- the gate being characterized is averaged in a such a way that the resulting process is depolarizing with the same average fidelity as the original gate. The depolarizing parameter of the averaged process is measured experimentally, and the result is related back to the average fidelity of the original gate. This technique is independent of the particular starting state of the experiment, and therefore is not affected by SPAM errors. However, RB has several shortcomings which make it unsatisfactory as a sole characterization technique for fault-tolerant QIP. For one thing, it is limited to Clifford gates\footnote{For recent work discussing generalizations of RB to non-Clifford gates, see Refs.~\cite{kimmel_robust_2014,dugas_characterizing_2015}.}, and so cannot be used to characterize a universal gate set for quantum computing. For another, RB provides only a single metric of gate quality, the average fidelity.\footnote{Recently, generalizations of RB have been proposed to measure leakage rates~\cite{wallman_robust_2015} and the coherence~\cite{wallman_estimating_2015} of errors. These advances have the promise to turn RB into a more widely applicable characterization tool.} This can be insufficient for determining the correct qubit error model to use for evaluating compatibility with QEC. Several groups have shown that qualitatively different errors can produce the same average gate fidelity, and in the case of coherent errors the depolarizing channel inferred from the RB gate fidelity underestimates the effect of the error~\cite{magesan_modeling_2013,gutierrez_approximation_2013,puzzuoli_tractable_2014}. 
Finally, RB assumes the errors on subsequent gates are independent. This assumption fails in the presence of non-Markovian, or time-dependent noise. GST suffers from this assumption as well, but the long sequences used in RB make this a more pressing issue.

Despite these apparent shortcomings, RB has been used with great success by several groups to measure gate fidelities and to diagnose and correct errors~\cite{gaebler_randomized_2012,barends_superconducting_2014,kelly_optimal_2014,muhonen_quantifying_2014}. RB also has the advantage of scalability -- the resources required to implement RB (number of experiments, processing time) scale polynomially with the number of qubits being characterized~\cite{magesan_scalable_2011,magesan_characterizing_2012}. QPT and GST, on the other hand, scale exponentially with the number of qubits. As a result, these techniques will foreseeably be limited to addressing no more than 2-3 qubits at a time. In our view, GST and RB will end up complementing each other as elements of a larger characterization protocol for any future multi-qubit quantum computer. We will not discuss RB further in this document.

This document reviews GST and provides instructions and examples for workers who would like to implement GST to characterize their qubits. The goal is to provide a guide that is both practical and self-contained. For simplicity, consideration is restricted to a single qubit throughout. We begin in Chapter~\ref{math-chapter} with a review of the mathematical background common to all characterization techniques. This includes the representation of gates as quantum maps via the process matrix and Pauli transfer matrix, the superoperator formalism, and the Choi-Jamiolkowski representation, which allows the maximum likelihood estimation (MLE) problem to be formulated as a semidefinite program (SDP). Chapter~\ref{derivation-chapter} begins with a review of quantum state and process tomography, the characterization techniques that underlie GST. We then continue to GST itself, including a derivation of linear-inversion gate set tomography (LGST) following Ref.~\cite{RBK2013}. Although LGST is weaker than MLE since it does not in general provide physical estimates, it is useful both as a starting point for the nonlinear optimization associated with MLE, and also in its own right as a technique for getting some information about the gate set quickly and with little numerical effort. This is followed by a description of the MLE problem in both the process matrix~\cite{Chow2009SI} and Pauli transfer matrix~\cite{Chow2012SI} formulations. MLE is the standard approach for obtaining physical gate estimates from tomographic data. In Chapter~\ref{implementation-chapter} we continue with GST, presenting the detailed experimental protocol as well as numerical results implementing the ideas of Chapter~\ref{derivation-chapter}. The implementation utilizes simulated data with simplified but realistic errors. This data was designed to incorporate the important properties of actual data including coherent and incoherent errors and finite sampling noise. Using this simulated data, we compare the performance of QPT and GST.

\chapter{Mathematical preliminaries}\label{math-chapter}
In this chapter we review the mathematical background on quantum operations and establish notation that we will use later on.

\section{Quantum operations}\label{quantops}

Quantum operations are linear maps on density operators $\rho \rightarrow \Lambda(\rho)$, that take an arbitrary initial state $\rho \in \mathcal{H}$ in some Hilbert space $\mathcal{H}$ and output another state $\Lambda(\rho)\in\mathcal{H}$. Although the formalism we present is applicable to arbitrary ($d$-level, or qudit) quantum systems, we will restrict ourselves to systems of qubits, i.e. $\mathcal{H} = \mathcal{H}_n$, where $\mathcal{H}_n$ is some $2^n$-dimensional (n-qubit) Hilbert space. This is done in the interest of concreteness, and because most experiments involve qubit systems. Thus, for example, we will explicitly use the familiar qubit basis of Pauli operators rather than a  more general qudit basis.

For such a map to describe a physical process, two basic requirements must be met: (1) for any initial state of the universe (system + environment), the final state after application of the map must have nonnegative probabilities for measuring the eigenstate of any observable (in other words, the density matrix of the universe is always positive semidefinite), and (2) total probability must be conserved.\footnote{Requirement (2) is relaxed in the presense of leakage errors (transitions outside the qubit subspace). In this case, the total probability must {\em not increase}. For an interesting discussion of non-trace preserving maps, see Ref.~\cite{bhandari_trace_preserving_2015}.} Requirement (1) is known as Complete Positivity (CP) and (2) is called Trace Preservation (TP), since the total probability of all eigenstates for the state $\rho$ is $\Tr\{\rho\}$. A physical map is then a CPTP map.

It turns out that a general CP map on just the system of interest (the environment having been traced out) can be written as~\cite{NielsenChuang}
\begin{equation}
\Lambda(\rho) = \sum_{i=1}^N K_i \rho K_i^\dag, \label{Kraus}
\end{equation}
for some $N\leq d^2$, where $d = 2^n$ is the Hilbert space dimension. This is called the {\em Kraus representation}. The $K_i$ are Kraus operators, and need not be unitary, hermitian, or invertible. The map is also TP if $\sum_i K_i^\dag K_i = I$, where $I$ denotes the identity.\footnote{To see this, we take the trace of Eq.~(\ref{Kraus}) and rearrange operators inside the trace: $\Tr\{\Lambda(\rho)\} = \Tr\{\sum_i K_i^\dag K_i \rho\}$. By trace preservation, $\Tr\{\Lambda(\rho)\} = \Tr\{\rho\}$. We therefore have $\Tr\{\sum_i K_i^\dag K_i \rho\} = \Tr\{\rho\}$ for any $\rho$, which can only be true if $\sum_i K_i^\dag K_i = I$.} This is called the {\em completeness condition}.

\subsection{Examples of quantum operations}

Eq. (\ref{Kraus}) contains unitary evolution as a special case: $N=1, K_1 \in SU(d)$. It also describes non-unitary processes, such as depolarization and amplitude damping, that can represent coupling to an environment (such as an external electromagnetic field). Below are some examples of the Kraus representation for a single qubit that describe familiar quantum processes.

\begin{enumerate}
\item {\em Depolarization} -- This operation replaces the input state with a completely mixed state with probability $p$, and does nothing with probability $1-p$. It describes a noise process where information is completely lost with probability $p$. For a single qubit, depolarization is given by the quantum map~\cite{NielsenChuang},
\begin{equation}
\Lambda_{dep}(\rho) = \left(1-\frac{3}{4}p\right)\rho + \frac{p}{4}\left(X\rho X + Y\rho Y + Z\rho Z\right). \label{example-depol}
\end{equation}
The Kraus operators are $\sqrt{1-3p/4}I,\sqrt{p}X/2,\sqrt{p}Y/2,\sqrt{p}Z/2$.\footnote{I, X, Y, Z are the single-qubit Pauli operators
\begin{eqnarray}
I = \left(\begin{array}{cc} 1 & 0 \\ 0 & 1 \end{array}\right),\; X = \left(\begin{array}{cc} 0 &1\\ 1 &0\end{array}\right),\; Y =\left(\begin{array}{cc} 0 & -i\\ i &0\end{array}\right),\; 
Z =\left(\begin{array}{cc} 1 & 0\\ 0 &-1\end{array}\right).\nonumber
\end{eqnarray}
}
\item {\em Dephasing} -- This process can arise when the energy splitting of a qubit fluctuates as a function of time due to coupling to the environment. Charge noise affecting a transmon qubit is of this type. Dephasing is represented by a phase-flip channel~\cite{NielsenChuang}, which describes the loss of phase information with probability $p$. This channel projects the state onto the $Z$-axis of the Bloch sphere with probability $p$, and does nothing with probability $1-p$:
\begin{equation}
\Lambda_{Z}(\rho) = \left(1-\frac{p}{2}\right)\rho + \frac{p}{2}Z\rho Z. \label{example-dephasing}
\end{equation}
Here the Kraus operators are $\sqrt{1-p/2}I,\sqrt{p/2}Z$. 
\item {\em Spontaneous Emission / Amplitude Damping} -- This process describes energy loss from a quantum system, the standard example being spontaneous emission from an atom~\cite{NielsenChuang}. The Kraus operators are
\begin{eqnarray}
K_0 = \left(\begin{array}{cc} 1 & 0 \\ 0 & \sqrt{1-p} \end{array}\right),\; K_1 = \left(\begin{array}{cc} 0 &\sqrt{p}\\ 0 &0\end{array}\right).\label{amp-damp-kraus-ops}
\end{eqnarray}
$K_1$ changes $|1\rangle$ to $|0\rangle$, corresponding to the emission of a photon to the environment. $K_0$ leaves $|0\rangle$ unchanged but reduces the amplitude of $|1\rangle$, representing the change in the relative probabilities of the two states when no photon loss is observed. Note that the form of $K_0$ is fixed by $K_1$, the completeness condition, and the requirement that $K_0$ not change the amplitude of the $|0\rangle$ state. 
\end{enumerate}

It is worth noting that at no point did we explicitly invoke the time dependence of quantum maps written in terms of Kraus operators. In fact, the time dependence is implicitly built in, through the parameter $p$ appearing in the examples above. Starting from Eq.~(\ref{Kraus}), and under the assumptions of local, Markovian noise, we can derive the Lindblad equation for the time evolution of the system density matrix, $\rho$. The Lindblad operators are related to the Kraus operators via $K_i = \sqrt{t} L_i$. (See for example, Ref.~\cite{Haroche-Raimond} Eq.~(4.62).) As a result, $p\propto t$, as we would expect from Fermi's golden rule for transitions induced by coupling to an environment. We do not worry about explicit time evolution in this document, and therefore use Eq.~(\ref{Kraus}) rather than the Lindblad equation, assuming implicitly that quantum operations take a finite time $t$.

For a given quantum operation, the Kraus operators are not unique~\cite{NielsenChuang}. They can be fixed by expressing them in a particular basis of $\mathcal{H}_n$. A typical choice is the Pauli basis, $\mathcal{P}^{\otimes n}$, where $\mathcal{P} = \{I,X,Y,Z\}$. We follow this convention here. 

\subsection{Process matrix}\label{process-matrix-section}
The Pauli basis leads to a useful representation of a quantum map: the process matrix, or $\chi$-matrix. Expanding $K_i$ in terms of Pauli operators, we obtain $K_i = \sum_{j=1}^{d^2}a_{ij}P_j$, $P_j \in \mathcal{P}^{\otimes n}$. Inserting this expression into Eq. (\ref{Kraus}) gives
\begin{equation}
\Lambda(\rho) = \sum_{j,k=1}^{d^2} \chi_{jk}P_j \rho P_k, \label{chi}
\end{equation}
where $\chi_{jk} = \sum_i a_{ij}a_{ik}^*$. $\chi$ is a $d^2 \times d^2$ complex-valued matrix, $d = 2^n$. The $\chi$-matrix completely determines the map $\Lambda$. From its definition in terms of the $a$-coefficients, we find that $\chi$ is Hermitian and positive semidefinite.\footnote{Indeed, $\chi_{kj} = \sum_i a_{ik}a_{ij}^* = \chi_{jk}^*$, so $\chi$ is Hermitian. Also, for any $d^2$-dimensional complex vector $\vec{v}$, we have $\vec{v}^\dag \chi \vec{v} = \sum_{jk} v_j^* \chi_{jk} v_k = \sum_{ijk} (a_{ij}v_j^*)(a_{ik}^*v_k) = \sum_i |\sum_j a_{ij}v_j^*|^2 \geq 0$. Therefore $\chi$ is positive-semidefinite.} In addition, we have $\sum_i K_i^\dag K_i = \sum_{ijk}a_{ik}^*P_k a_{ij} P_j = \sum_{jk} (\sum_i a_{ij}a_{ik}^*)P_k P_j = \sum_{jk} \chi_{jk}P_kP_j$. Thus for a TP map, the completeness condition reads $\sum_{jk} \chi_{jk}P_kP_j = I$. Hence, the $\chi$-matrix for a physical (CPTP) map has $d^2 (d^2-1)$ free parameters. (A $d^2 \times d^2$ complex hermitian matrix has $d^4$ free parameters, and the completeness condition adds $d^2$ constraints.)

\subsubsection{Examples}

\begin{itemize}
\item[1.] {\em Depolarization} -- The depolarizing channel, Eq.~(\ref{example-depol}), is already written in the form of Eq.~(\ref{chi}). We can simply read off the coefficients of the process matrix,
\begin{eqnarray}
\chi = \left(\begin{array}{cccc} 1-3p/4 & 0 & 0 & 0\\ 0 & p/4 & 0 & 0\\ 0 & 0 & p/4 & 0\\ 0 & 0 & 0 & p/4 \end{array}\right).\nonumber
\end{eqnarray}
\item[2.] {\em Dephasing} -- As for depolarization, Eq.~(\ref{example-dephasing}) is already written in the form of Eq.~(\ref{chi}). The process matrix is,
\begin{eqnarray}
\chi = \left(\begin{array}{cccc} 1-p/2 & 0 & 0 & 0\\ 0 & 0 & 0 & 0\\ 0 & 0 & 0 & 0\\ 0 & 0 & 0 & p/2 \end{array}\right).\nonumber
\end{eqnarray}
\item[3.] {\em Rotations} -- A rotation by angle $\theta$ about axis $\hat{n}$ applied to a state $\rho$ is given by $\Lambda_\theta(\rho) = e^{-i\theta \hat{n}\cdot\hat{\sigma}/2}\rho e^{i\theta \hat{n}\cdot\hat{\sigma}/2}$. For simplicity, let us choose the rotation axis to be the z-axis. Expanding $e^{-i\theta \hat{n}\cdot\hat{\sigma}/2} = \cos(\theta/2)-i\hat{n}\cdot\hat{\sigma}\sin(\theta/2)$, and setting $\hat{n} = \hat{z}$, we get
\begin{eqnarray}
\chi = \frac{1}{2}\left(\begin{array}{cccc} 1+\cos(\theta) & 0 & 0 & i\sin(\theta)\\ 0 & 0 & 0 & 0\\ 0 & 0 & 0 & 0\\  -i\sin(\theta) & 0 & 0 & 1-\cos(\theta) \end{array}\right).\nonumber
\end{eqnarray}
\item[4.] {\em Spontaneous Emission / Amplitude Damping} -- Expanding the Kraus operators for amplitude damping, Eq.~(\ref{amp-damp-kraus-ops}), in terms of the Paulis, we find
\begin{eqnarray}
\chi = \frac{1}{2}\left(\begin{array}{cccc} (1+\sqrt{1-p})^2 & 0 & 0 & p/2\\ 0 & p/2 & -ip/2 & 0\\ 0 & ip/2 & p/2 & 0\\  p/2 & 0 & 0 & (1-\sqrt{1-p})^2 \end{array}\right).\nonumber
\end{eqnarray}
\end{itemize}

\subsection{Pauli transfer matrix}\label{ptm-section}

Another useful representation of a quantum map  is the Pauli transfer matrix (PTM):
\begin{equation}
(R_\Lambda)_{ij} = \frac{1}{d}\Tr\{P_i \Lambda(P_j)\}. \label{PTM}
\end{equation}
The PTM has several convenient properties. Since $\Lambda(\rho)$ is a density matrix, it can be expanded in the $\{P_i\}$ with real coefficients in the interval [-1,1]. (This is true for any Hermitian operator.) Inspection of the right-hand side of Eq.~(\ref{PTM}) then shows that all entries of the PTM must also be real and in the interval [-1,1]. Also, as we show in Section~\ref{superoperators}, the PTM of a composite map is simply the matrix product of the individual PTMs. This makes it easy to evaluate the result of multiple gates acting in succession. The process matrix does not have this property. We note that $R$ is generally not symmetric.

In the PTM representation, the trace preservation (TP) constraint has a particularly simple form. Since $\Tr\{P_j\}=\delta_{0j}$, we must have $\Tr\{\Lambda(P_j)\} = \delta_{0j}$. But $\Tr\{\Lambda(P_j)\} = \Tr\{P_0\Lambda(P_j)\} = (R_\Lambda)_{0j}$. Therefore the TP constraint is $(R_\Lambda)_{0j} = \delta_{0j}$. In other words, the first row of the PTM is one and all zeros.

Compete positivity on the other hand is not expressed as simply in the PTM representation as it is for the $\chi$-matrix. We discuss this more in Sec.~\ref{physicality-section}.

Another sometimes useful condition that is conveniently expressible via the PTM is {\em unitality}. A unital map is one that takes the identity to the identity. Physically, this is a map that does not increase the purity of a state, i.e. it does not `unmix' a mixed state. Many quantum operations of interest, including unitary maps, are unital. One interesting exception is amplitude damping, as we will see in the examples below. Mathematically, unitality is expressed as $\Lambda(I) = I$. Inserting this into the definition of the PTM, Eq.~(\ref{PTM}), we find $(R_\Lambda)_{i0} = \Tr\{P_i \Lambda(I)\} = \Tr\{P_i I\} = \delta_{i0}$. Therefore the PTM for a unital map must have one and all zeros in its first column.

The PTM and process matrix representations are equivalent, and one can express one in terms of the other. Substituting Eq. (\ref{chi}) into Eq. (\ref{PTM}) with $\rho = P_j$ gives $(R_\Lambda)_{ij} = \frac{1}{d}\sum_{kl}\chi_{kl}\Tr\{P_i P_k P_j P_l\}$. (Note that $\rho = P_j$ is not a physical density operator: $\Tr\{P_j\} \neq 1$. But $\Lambda(P_j)$ is still formally defined.) The inverse transformation (expressing $\chi$ in terms of $R$) cannot be written as a compact analytical expression but can be implemented straightforwardly for any particular PTM using symbolic or numerical software.

\subsubsection{Examples}

Using the above definitions, we can write down the PTMs for the examples in Sec.~\ref{process-matrix-section}. (See that section for definitions of $p$ and $\theta$.)

\begin{itemize}
\item[1.] {\em Depolarization}
\begin{eqnarray}
R = \left(\begin{array}{cccc} 1 & 0 & 0 & 0\\ 0 & 1-p & 0 & 0\\ 0 & 0 & 1-p & 0\\ 0 & 0 & 0 & 1-p \end{array}\right).\nonumber
\end{eqnarray}
\item[2.] {\em Dephasing} 
\begin{eqnarray}
R = \left(\begin{array}{cccc} 1 & 0 & 0 & 0\\ 0 & 1- p & 0 & 0\\ 0 & 0 & 1 - p & 0\\ 0 & 0 & 0 & 1 \end{array}\right).\nonumber
\end{eqnarray}
\item[3.] {\em Rotation by $\theta$ about z-axis}
\begin{eqnarray}
R = \left(\begin{array}{cccc} 1 & 0 & 0 & 0\\ 0 & \cos(\theta) & -\sin(\theta) & 0\\ 0 & \sin(\theta) & \cos(\theta) & 0\\ 0 & 0 & 0 & 1 \end{array}\right).\nonumber
\end{eqnarray}
\item[4.] {\em Spontaneous Emission / Amplitude Damping}
\begin{eqnarray}
R = \left(\begin{array}{cccc} 1 & 0 & 0 & 0\\ 0 & \sqrt{1-p} & 0 & 0\\ 0 & 0 & \sqrt{1-p} & 0\\ p & 0 & 0 & 1-p \end{array}\right).\nonumber
\end{eqnarray}
\end{itemize}

These examples illustrate the convenient structure of the PTM. Note in particular examples 3 and 4. For rotations, the PTM reduces to block-diagonal form consisting of the 1-dimensional identity matrix and a $3 \times 3$ rotation matrix. This rotation matrix determines the transformation of the 1-qubit Bloch vector under the rotation map. For amplitude damping, the form of the PTM is much simpler than that of the corresponding $\chi$-matrix and makes the non-unitality apparent.

From the PTMs, we can see by inspection that depolarization, dephasing, and rotations are all trace preserving and unital. Amplitude damping, however, is trace preserving but not unital (the first column of the PTM has a non-zero entry beyond the first entry). Physically, amplitude damping corresponds to spontaneous emission from the $|1\rangle$ to the $|0\rangle$ state. Therefore, any mixed state tends towards the pure state $|0\rangle$ at long enough times. (This is in fact the procedure typically used experimentally to ``reset" a qubit to the $|0\rangle$ state -- waiting long enough for spontaneous emission to occur.)

We note in passing that invertible quantum maps form a group. It is therefore natural to look for group representations, since this allows quantum maps to be represented as matrices, and map composition as the product of representation matrices. Such group representations are called superoperators. We saw above that PTMs in fact have the properties we just described. Therefore PTMs form a superoperator group representation, based on the Pauli matrices. We study this representation below, and show how it can be used to simplify the description of quantum maps.

\section{Superoperator formalism}\label{superoperators}

Manipulations with quantum maps can be conveniently carried out using the superoperator formalism~\cite{KitaevBook}. In this formalism, density operators $\rho$ on a Hilbert space of dimension $d$ are represented as vectors $|\rho\rb$ in {\em Hilbert-Schmidt} space of dimension $d^2$. Quantum operations (linear maps on density operators) are represented as matrices of dimension $d^2 \times d^2$. The Hilbert-Schmidt inner product is defined as $\lb A| B\rb = \Tr\{A^\dag B\}/d$, where $A$, $B$ are density operators. As we shall see, defining the inner product in this way enables map composition to be represented as matrix multiplication. It also allows us to make contact with measurements. The expected value $p$ of a POVM $E$ for a state $\rho$ is $p = \Tr\{E\rho\}$.\footnote{POVM stands for Positive Operator-Valued Measure, and denotes a formalism for describing measurements. A POVM $E$ is also called a measurement operator and is used to represent the average outcome of measuring a state $\rho$, through the equation $p = \Tr\{E\rho\}$. POVMs are hermitian, positive semidefinite, and satisfy the completeness relation $\sum_i E_i = I$, where $i$ enumerates the possible experimental outcomes. In this document we are concerned primarily with the special case of projective measurements, for example $E=|0\rangle\langle 0|$, and use the POVM formalism simply as a matter of notational convenience. For this reason, we do not discuss POVMs in any detail, and refer the reader to Ref.~\cite{NielsenChuang}, Ch. 2.2.6 for a complete treatment.}

Although superoperators can be defined relative to any basis of $\mathcal{H}_n$, we continue to find it convenient to use the Pauli basis introduced above. From this point on, we use rescaled Pauli operators $P_i \rightarrow P_i/\sqrt{d}$. This way the basis is properly normalized, and we avoid having to write factors of $d$ everywhere.

We represent states $\rho$ as state vectors $|\rho\rb = \sum_k |k\rb\lb k|\rho\rb$, with components
\begin{eqnarray}
\lb k | \rho \rb &=& \Tr\{P_k \rho\}, \label{superop-component}\\
P_k &\in& \{I/\sqrt{2},X/\sqrt{2},Y/\sqrt{2},Z/\sqrt{2}\}^{\otimes n}. \label{PauliBasis}
\end{eqnarray}
This is simply a restatement of the completeness of the Pauli basis: $\rho = \sum_k P_k \Tr\{P_k \rho\}$ for any $\rho$, which implies that the operator $\sum_k P_k \Tr\{P_k \,\Cdot\,\}$ equals the identity. In superoperator notation, this is $\sum_k |k\rb\lb k| = I$, from which Eq.~(\ref{superop-component}) follows.

Quantum operations are represented as matrices, 
\begin{eqnarray}
R_\Lambda &=& \sum_{jk} |j\rb\lb j| R_\Lambda |k\rb\lb k|, \\
\lb j| R_\Lambda |k\rb &=& \Tr\{P_j \Lambda (P_k)\}.\label{superop-PTM}
\end{eqnarray} 
Note that $\lb j| R_\Lambda |k\rb$ is just the PTM, $(R_\Lambda)_{jk}$. (The $1/d$ factor was suppressed by our rescaling.)

From these definitions, it is straightforward to derive the following consequences:
\begin{eqnarray}
|\Lambda(\rho)\rb &=& R_\Lambda|\rho\rb,\label{superop-vec}\\
R_{\Lambda_2 \circ \Lambda_1} &=& R_{\Lambda_2} R_{\Lambda_1}.\label{superop-prod} 
\end{eqnarray}

\begin{proof} To prove Eq.~(\ref{superop-vec}), we first use the completeness relation and then Eq.~(\ref{superop-component}),
\begin{equation}
|\Lambda(\rho)\rb = \sum_k |k\rb\lb k|\Lambda(\rho)\rb = \sum_k |k\rb \Tr\{P_k \Lambda(\rho)\}.\nonumber
\end{equation}
Next, we expand $\rho = \sum_j P_j \Tr\{P_j \rho\}$ and insert into the right-hand-side of the above equation, giving
\begin{equation}
|\Lambda(\rho)\rb = \sum_{jk} |k\rb \Tr\{P_k \Lambda(P_j)\}\Tr\{P_j \rho\}.\nonumber
\end{equation}
Substituting $\Tr\{P_k \Lambda(P_j)\} = \lb k | R_\Lambda | j\rb$ (see Eq.~(\ref{superop-PTM})) and $\Tr\{P_j \rho\} = \lb j|\rho\rb$ (Eq.~(\ref{superop-component}) again) we obtain
\begin{equation}
|\Lambda(\rho)\rb = \sum_{jk} |k\rb \lb k|R_\Lambda|j\rb\lb j|\rho\rb.\nonumber
\end{equation}
Finally, we eliminate the sums over $j, k$ using the completeness relation, thus proving Eq.~(\ref{superop-vec}).

To prove Eq.~(\ref{superop-prod}), we act with $R_{\Lambda_2 \circ \Lambda_1}$ on an arbitrary state $|\rho\rb$, and use Eq.~(\ref{superop-vec}).
\begin{equation}
R_{\Lambda_2 \circ \Lambda_1}|\rho\rb = |\Lambda_2\left(\Lambda_1(\rho)\right)\rb.\nonumber
\end{equation}
Using Eq.~(\ref{superop-vec}) again, we find
\begin{equation}
|\Lambda_2\left(\Lambda_1(\rho)\right)\rb = R_{\Lambda_2}|\Lambda_1(\rho)\rb =  R_{\Lambda_2}R_{\Lambda_1}|\rho\rb.\nonumber
\end{equation}
This proves Eq.~(\ref{superop-prod}).
\end{proof}

In this way, the composition of quantum maps has been expressed as matrix multiplication. We now have a simple way to write down experimental outcomes. Consider the following experiment: prepare state $\rho$, perform quantum operation $\Lambda$, measure the POVM $E$. Repeat many times and calculate the average measured value, $m$. The expected value, $p = \mathrm{E}(m)$, of the measurement $m$ is the quantum expectation value, which can be written as $p=\Tr\{E\Lambda(\rho)\} = \lb E | R_\Lambda |\rho\rb$.

\section{Physicality constraints}\label{physicality-section}

As mentioned in Sec.~\ref{quantops}, a physical map must be completely positive and trace preserving (CPTP), which is equivalent to the following two conditions: (1) for any initial state of the universe, the final state after application of the map must have nonnegative probabilities for measuring the eigenstate of any observable (in other words, the density matrix of the universe is always positive semidefinite), and (2) total probability must be conserved. The mathematical statement of these requirements is different for the process matrix and PTM representations of a quantum map, and will be used later on to write down the constraints for the optimization problem for the gate set.

\subsection{Process matrix representation}

We have seen that hermiticity and positive semidefiniteness of the $\chi$ matrix follow from the Kraus representation, which (as we stated without proof) is a consequence of complete positivity. Also, we used the Kraus representation to show that trace preservation implies the completeness condition, $\sum_{ij} \chi_{ij}P_j P_i = I$. Since the process matrix is so central to our work, it is interesting to see how these constraints follow directly from the CPTP requirement and Eq.~(\ref{chi}), which is a general way of writing any quantum map. We now show this.

Since the CP condition must hold for any initial state, we consider initial states of the form $\rho = |\phi\rangle\langle\phi|\otimes I$, where $|\phi\rangle$ is a pure state of the system, or qubit, and $I$ refers to the ``rest of the universe''. We can neglect reference to the ``rest of the universe" since it will be traced out of all formulas in which the map $\Lambda$ acts only on the system. Therefore we omit the $\otimes I$ below.

The probability of observing the pure state $|\psi\rangle$ of the system after application of the map $\Lambda$ to the initial state $|\phi\rangle$ is
\begin{equation}
p_\psi = \Tr\{\Lambda(|\phi\rangle\langle\phi|)|\psi\rangle\langle \psi|\} = \langle \psi | \Lambda(|\phi\rangle\langle\phi|) |\psi\rangle.\label{pk-proof2}
\end{equation}
Using Eq.~(\ref{chi}), we obtain
\begin{equation}
\Lambda(|\phi\rangle\langle\phi|) = \sum_{ij}\chi_{ij}P_i|\phi\rangle\langle \phi| P_j.
\end{equation}
Inserting this into Eq.~(\ref{pk-proof2}) we get
\begin{equation}
p_\psi =  \sum_{ij}\chi_{ij}\langle \psi|P_i|\phi\rangle\langle \phi| P_j|\psi \rangle.\label{pk2-proof2}
\end{equation}
Since $p_\psi$ is real, we find upon setting $p_\psi^* = p_\psi$ and interchanging the dummy indices $i,j$ in Eq.~(\ref{pk2-proof2}), that $\chi^\dag = \chi$.

Next, let us define a vector $\vec{v}$ with components $v_j = \langle \phi| P_j|\psi \rangle$. Then Eq.~(\ref{pk2-proof2}) becomes $p_\psi = \vec{v}^\dag \chi \vec{v}$. Since the probability $p_\psi \geq 0$ and $|\psi\rangle$, $|\phi\rangle$ are arbitrary, the matrix $\chi$ must be positive semidefinite.

Finally, the completeness condition is proved just like for the Kraus representation. Trace preservation means $\Tr\{\Lambda(\rho)\} = \sum_{ij}\chi_{ij}\Tr\{P_jP_i\rho\} = \Tr\{\rho\}$. Since this must hold for any $\rho$, the completeness condition follows.

Although we have only proved that these conditions on $\chi$ are necessary for CPTP (we only considered a restricted set of initial states), it turns out they are also sufficient. We use this in the formulation of the constraints for the optimization problem in Ch.~\ref{derivation-chapter}. 

It is also clear now that the Kraus representation can be viewed as a special case of a general $\chi$-representation, where the process matrix is the identity. Indeed, the CP constraint implies that $\chi$ must have real and positive eigenvalues. Starting from Eq.~(\ref{chi}) for an arbitrary (not necessarily Pauli) basis $P_i$, we can diagonalize $\chi$ and absorb the (real and positive) eigenvalues into the definition of the $K_i$, resulting in Eq.~(\ref{Kraus}).

\subsection{PTM representation} \label{ptm-physicality}

As we saw above, the TP constraint in the PTM representation is  $(R_\Lambda)_{0j} = \delta_{0j}$. The CP constraint, however, cannot be expressed in terms of the PTM directly. Instead, one typically uses the fact~\cite{magesan_gate_2011} that the  Choi-Jamiolkowski (CJ) matrix~\cite{choi_completely_1975,jamiolkowski_linear_1972} associated with a CP map is positive semidefinite. The CJ matrix is a density matrix - like representation of a quantum map that lives in a tensor product of two Hilbert-Schmidt spaces. For a quantum map $\Lambda$ it can be written in terms of the PTM as~\cite{Merkel2013}
\begin{equation}
\rho_\Lambda = \frac{1}{d^2}\sum_{i,j=1}^{d^2} (R_\Lambda)_{ij} P_j^T \otimes P_i. \label{choi1} 
\end{equation}
In terms of the CJ matrix, the CP constraint is $\rho_\Lambda \succcurlyeq 0$, where the curly inequality denotes positive-semidefiniteness. See Ref.~\cite{magesan_gate_2011} for further discussion.

\chapter{Derivation of GST}\label{derivation-chapter}

In this chapter, we derive GST and describe some of the relevant data analysis techniques. In the following chapter, we present the experimental and data analysis protocols in step-by-step fashion. The reader interested in getting right to the implementation may want to skip ahead to that chapter.

GST can be viewed as a self-consistent extension of quantum process tomography (QPT), which itself grew out of quantum state tomography (QST). We therefore begin with a brief review of QST and QPT.

\section{Quantum state tomography}\label{qst-section}

Quantum state tomography~\cite{leibfried_experimental_1996} attempts to characterize an unknown state $\rho$ by measuring its components $\lb k |\rho\rb$, usually in the Pauli basis. It can be more convenient to use a different basis of measurement operators $E_j$, $j=1,\ldots,d^2$ that span the Hilbert-Schmidt space $\mathcal{B}(\mathcal{H}_d)$.\footnote{We here use the notation $\mathcal{H}_d$ instead of $\mathcal{H}_n$ as in Ch.~\ref{math-chapter} to keep the discussion momentarily more general. For $n$ qubits, $d=2^n$.} Generally, one measures the $d^2$ {\em probabilities}
\begin{equation}
p_j = \lb E_j|\rho\rb = \sum_k \lb E_j|k\rb\lb k|\rho\rb.\label{p-qst}
\end{equation}
The matrix $A_{jk}=\lb E_j|k\rb$ is assumed known, since the $\lb E_j|$ were chosen in advance by the experimenter, and the $|k\rb$ are Pauli basis vectors.

The probability $p_j$ can be estimated as the sample average $m_j= \sum_{i=1}^N m_{ij}/N $ of $N$ single-shot measurements of $E_j$, where $m_{ij} = 0$ or 1 is the outcome of the $i$-th measurement. Since $p_j$ is the true probability, the expected value of $m_j$ is ${\rm E}(m_j) = \sum_{i=1}^N {\rm E}(m_{ij})/N = p_j$, and the variance is ${\rm Var}(m_j) = p_j(1-p_j)/N$. The sample average $m_j$ approaches the true value $p_j$ as $N\rightarrow \infty$. 

Since $A_{jk}$ is known, Eq.~(\ref{p-qst}) can be solved by matrix inversion, $|\hat{\rho}\rb = A^{-1}|m\rb$, where $|m\rb$ is the vector of measurements $m_j$ and $\hat{\rho}$ is the linear inversion estimator for $\rho$. The only requirement is that $E_j$ be chosen such that $A$ is invertible.

Estimation is particularly simple if $|E_j\rb =|j\rb$. Then no matrix inversion is necessary and $|\hat{\rho}\rb = |m\rb$. (Technically $m_0$ should be set to $1/\sqrt{d}$, so that $\Tr\{P_0 \rho\}=1$.)

\section{Quantum process tomography}\label{qpt-section}

Quantum process tomography~\cite{chuang_prescription_1997,poyatos_complete_1997} is similar to QST, but now the goal is to characterize a gate $G$ rather than a state. This is done by measuring the process matrix $\chi$ or the PTM, $\lb k|G|l\rb$, via measurements of the $d^4$ probabilities
\begin{equation}
p_{ij} = \lb E_j|G|\rho_i\rb.\label{p-qpt}
\end{equation}
Here the $E_j$ are as before, and the set of vectors $|\rho_i\rb$ is a complete basis for the Hilbert-Schmidt space. Since we must have $\Tr\{\rho\} = 1$ and the Pauli matrices are traceless, we cannot choose $|\rho_i\rb = |i\rb$. Therefore, each $|\rho_i\rb$ must be chosen as some linear combination of Pauli basis vectors that can be inverted to give the PTM. Inserting complete sets of states in Eq.~(\ref{p-qpt}),
\begin{equation}
p_{ij} = \sum_{kl} \lb E_j|k\rb\lb k|G|l\rb \lb l|\rho_i\rb.\label{p-qst2}
\end{equation}
As before, the vectors $\lb E_j|$, $|\rho_i\rb$ are specified by the experimenter, and so the matrices $\lb E_j|k\rb$, $ \lb l|\rho_i\rb$ are assumed known. We can arrange the product of these two matrices into a single matrix, $S_{i+(j-1)d^2;k+(l-1)d^2} = \lb E_j|k\rb \lb l|\rho_i\rb$. Also, we can vectorize the PTM, defining $r_{k+(l-1)d^2} = G_{kl}$. Then, Eq.~(\ref{p-qst2}) becomes
\begin{equation}
\vec{p} = S\vec{r}.
\end{equation}
Given a vector of measurements $\vec{m}$ of the probabilities $\vec{p}$, this equation can be inverted to yield an estimate of the PTM:
\begin{equation}
\hat{\vec{r}} = S^{-1}\vec{m}.
\end{equation}

In the discussion so far, we have assumed S is full-rank. ($\{E_j\}$, $\{\rho_i\}$ each form a complete basis, and S is $d^4 \times d^4$.) The estimation can be improved by including an overcomplete ($>d^2$) set of states and measurements. Then $S$ has more rows than columns and cannot be inverted, but $S^\dag S$ can be. In this case, we obtain the ordinary least-squares estimate as
\begin{equation}
\hat{\vec{r}} = (S^\dag S)^{-1}S^\dag\vec{m}. \label{qpt-ols-estimate}
\end{equation}

In practice QST and QPT estimates are obtained using maximum likelihood estimation rather than linear inversion, since this allows physicality constraints to be put in. We discuss MLE in the context of GST in Sec.~\ref{mle-section} below.

\section{Gate set tomography}

We have seen that QST and QPT assume the initial states and final measurements are known. In fact, these states must be prepared using quantum gates which themselves may be faulty: $\lb E_j| = \lb E|F_j$, $|\rho_i\rb = F'_i|\rho\rb$, where $\lb E|$, $|\rho\rb$ are some particular starting state and measurement the experimenter is able to implement. This results in a self-consistency problem. If the state preparation and measurement (SPAM) gates $\{F_j\}$, $\{F'_i\}$ are sufficiently faulty, the QST and QPT estimates will be faulty as well. GST solves this problem by including the SPAM gates self-consistently in the gate set to be estimated.

The goal of GST is to completely characterize an unknown set of gates and states~\cite{RBK2013},
\begin{equation}
\mathcal{G} = \{|\rho\rb,\lb E|,G_0, ..., G_K\}, \label{gateset}
\end{equation}
where  $|\rho\rb \in \mathcal{H}_n$ is some initial state, $\lb E| \in \mathcal{H}_n^*$ is a 2-outcome POVM, and each $G_k \in \mathcal{B}(\mathcal{H}_n)$ is an n-qubit quantum operation in the Hilbert-Schmidt space $\mathcal{B}$ of linear operators on $\mathcal{H}_n$. $\mathcal{G}$ is called the {\em gate set}. By {\em characterization} we mean a procedure to estimate the complete process matrices or PTMs for $\mathcal{G}$ based on experimental data. Sometimes we would like to focus only on the gates and not the states and measurements, and we write the gate set as $\mathcal{G}=\{G_0,\ldots,G_K\}$. The distinction should be clear from the context.

As in QPT, the information needed to reconstruct each gate $G_k$ is contained in measurements of $\lb E_i |G_k|\rho_j\rb$, the gate of interest sandwiched between a complete set of states and POVMs. The experimental requirements for GST are therefore similar to QPT: the ability to measure expectation values of the form $p=\Tr\{E G(\rho)\} = \lb E | R_G |\rho\rb$ for the gate set, Eq. (\ref{gateset}). 

In QPT, the set $\{\lb E_i|, |\rho_j\rb\}$ is assumed given, and this leads to incorrect estimates when there are systematic errors on the gates preparing the initial and final states. For self-consistency, we must treat these SPAM gates on the same footing as the original gates $\{G_k\}$. This is done by introducing the SPAM gates\footnote{Reference~\cite{RBK2013} refers to these gates as {\em fiducial} gates, hence the letter $F$. We follow reference~\cite{Merkel2013} and refer to them as SPAM gates in order to make contact with the notion of SPAM introduced earlier.} explicitly: $|\rho_j\rb = F_j|\rho\rb$ and $\lb E_i| = \lb E |F_i$. The SPAM gates $\mathcal{F} = \{F_1,\ldots,F_N\}$ are composed of gates in the gate set $\mathcal{G}$, and therefore the minimal $\mathcal{G}$ must include sufficient gates to create a complete set of states and measurements. In contrast to QPT, it is not possible in GST to characterize one gate at a time. Instead, GST estimates every gate in the gate set simultaneously.\footnote{More accurately, there is a minimal set of gates that must be estimated all at once, which includes the gates required for SPAM. Additional gates may be added one at a time.}

Because the SPAM gates are included in the gate set, GST requires that only a single initial state $\rho$ be prepared and a single measurement $E$ implemented. This is close to the experimental reality, where $\rho$ is usually the ground state of a qubit or set of qubits, and $E$ is a measurement in the $Z$-basis.

\section{Linear inversion GST}\label{derivation-lgst}

We now derive a simple, closed-form algorithm for obtaining self-consistent gate estimates. This algorithm was introduced by Robin Blume-Kohout, et al.~\cite{RBK2013} and was inspired by the Gram matrix methods of Cyril Stark~\cite{stark_self-consistent_2014}. The limitation of linear-inversion GST (or LGST) is that it does not constrain the estimates to be physical. There is often an unphysical gate set that is a better fit to the data than any physical one. As a result, LGST by itself is insufficient to provide gate quality metrics.
Nevertheless, LGST provides a convenient method for diagnosing gate errors, and also gives a good starting point for the constrained maximum likelihood estimation (MLE) approaches we discuss later. Also, in cases where the LGST estimate happens to be physical, it is identical to the one found by MLE.

We begin by identifying a set of SPAM gate strings $\mathcal{F} = \{F_1,...,F_{d^2}\}$ that, when applied to our unknown fixed state $|\rho\rb$ and measurement $\lb E|$, produce a complete set of initial states $|\rho_k\rb = F_k |\rho\rb$, and final states $\lb E_k| = \lb E| F_k$.\footnote{Note that in defining $\lb E_k|$ we use $F_k$ and not $F_k^\dag$ as one might expect. This is important in the analysis that follows.} In terms of the Pauli basis, Eq. (\ref{PauliBasis}), the components of these states are
\begin{eqnarray}
|\rho_k\rb_i &=& \lb i|F_k|\rho\rb = \Tr\{P_i F_k(\rho)\}, \label{rho-component}\\
\lb E_k|_i &=& \lb E | F_k |i\rb = \Tr\{F_k^\dag(E) P_i\}. \label{E-component}
\end{eqnarray}

The SPAM gates must be composed of members from our gate set, $\mathcal{G}$.\footnote{If $\mathcal{G}$ is insufficient to produce a complete set of states and measurements in this way, we must add gates to $\mathcal{G}$ until this is possible. In practical applications, one is interested in characterizing a complete set of gates, so this is not a problem.} (For a single qubit, the completeness requirement means that $\{F_k |\rho\rb\}_{k=1}^{d^2}$ must span the Bloch sphere, and similarly for $\{\lb E | F_k\}$.) In general, SPAM gates have the form~\cite{RBK2013},
\begin{equation}
F_k = G_{f_{k1}}\circ G_{f_{k2}} \circ ... G_{f_{kL_k}}, \label{SPAMgates}
\end{equation}
where $\{f_{kl}\}$ are indices labeling the gates of $\mathcal{G} = \{G_0,...,G_K\}$, and $L_k$ is the length of the k-th SPAM gate string. 

\subsection{Example gate sets}\label{example}
It is useful to have concrete examples of gate sets. Here we consider two of the simplest, both for a single qubit. We will continue to use both examples below. (We use the notation $A_\theta$ to denote a rotation by angle $\theta$ about axis $A$, for example $X_{\pi/2}$ is a $\pi/2$ rotation about the $X$-axis of the Bloch sphere.)

\begin{itemize}
\item[1.] $\mathcal{G} = \{\{\},X_{\pi/2},Y_{\pi/2}\} = \{G_0,G_1,G_2\}$. The symbol $\{\}$ denotes the ``null'' gate -- do nothing for no time. (We will always choose $G_0$ to be the null gate for reasons that will become clear later.) One choice of SPAM gates is $\mathcal{F} = \{\{\},X_{\pi/2},Y_{\pi/2},X_{\pi/2}\circ X_{\pi/2} \} = \{G_0,G_1,G_2,G_1\circ G_1\}$. It is easy to see that the set of states $F_k |\rho\rb$ (measurements $\lb E| F_k)$ spans the Bloch sphere for any pure state $|\rho\rb$ (measurement $\lb E|$). 

\item[2.] $\mathcal{G} = \{\{\},X_{\pi/2},Y_{\pi/2},X_{\pi}\} = \{G_0,G_1,G_2,G_3\}$. Here the SPAM gates are $\mathcal{F} = \{\{\},X_{\pi/2},Y_{\pi/2},X_{\pi} \} = \mathcal{G}$. This includes one more gate in the gate set compared to Example 1, but now the SPAM gates do not contain products of gates from the gate set.
\end{itemize}

In Example 1, we chose $F_4 = X_{\pi/2}\circ X_{\pi/2}$ rather than $X_\pi$ so that we would not need to add the additional gate $X_\pi$ to the gate set $\mathcal{G}$. The choice $F_4 = G_1\circ G_1$ is more efficient from an experimental standpoint. This can be a consideration if the experimental resources required to implement GST for an additional gate (16 additional experiments) are significant. If this is not a consideration, it turns out to be a disadvantage for the purposes of data analysis to have multiples of the same gate within a single SPAM gate. This is because the order (the polynomial power in the optimization parameters) of the MLE objective function is proportional to the number of gates in the experiment. (This is not so much of an issue if an LGST analysis is sufficient.) Therefore, the choice of gate set depends on the constraints of the experimental implementation as well as the post-processing requirements.

\subsection{Gram matrix and A, B matrices}\label{gram-ab-section}
In GST, we work with expectation values,
\begin{equation}
p_{ijk} = \lb E | F_i G_k F_j |\rho\rb, \label{pijk-0}
\end{equation}
where $F_i,F_j \in \mathcal{F}$ and $G_k \in \mathcal{G}$. These quantities correspond to measurements that can (in principle) be carried out in the lab. Inserting a complete set of states on each side of $G_k$ in Eq.~(\ref{pijk-0}), we obtain
\begin{equation}
p_{ijk} = \sum_{rs}\lb E | F_i|r\rb\lb r| G_k |s\rb\lb s| F_j |\rho\rb \equiv \sum_{rs}A_{ir}(G_k)_{rs}B_{sj}. \label{pijk}
\end{equation}
This defines a set of Pauli transfer matrices, as follows. $(G_k)_{rs}=\lb r| G_k |s\rb$ is the $rs$-component of the PTM for the gate $G_k$. $A_{ir}=\lb E | F_i|r\rb$ and $B_{sj}=\lb s| F_j |\rho\rb$ are, respectively, the $r$-component of $\lb E_i|$ and the $s$-component of $|\rho_j\rb$ (See Eqs~(\ref{rho-component}),~(\ref{E-component})). It is useful to write $A$ and $B$ in component-free notation as
\begin{eqnarray}
A &=& \sum_i |i\rb\lb E|F_i, \\
B &=& \sum_j F_j|\rho\rb\lb j|.
\end{eqnarray}
One can easily verify that $A_{ir}=\lb E | F_i|r\rb$ and $B_{sj}=\lb s| F_j |\rho\rb$ as required. We then find that, according to Eq.~(\ref{pijk}), $p_{ijk} = (AG_kB)_{ij}$.

Experimentally measuring the values $p_{ijk}$ in Eq. (\ref{pijk}) amounts to measuring the ($ij$) components of the matrix
\begin{equation}
\tilde{G}_k = AG_k B.
\end{equation}
Since $G_0 = \{\}$ (the null gate), the $k=0$ experiment gives 
\begin{equation}
g = \tilde{G}_0 = AB. \label{gram}
\end{equation}
The matrix $g$ is the Gram matrix of the $\{F_i\}$ in the Pauli basis.\footnote{The importance of the null gate is now clear. In order for Eq. (\ref{gram}) to hold, \{\} must be an {\em exact} identity. Anything else, e.g. an idle gate, would introduce an error term between $A$ and $B$.} We observe:
\begin{equation}
g^{-1}\tilde{G}_k = B^{-1}A^{-1}AG_kB = B^{-1}G_kB.\label{g-inv-G}
\end{equation}
Thus,
\begin{equation}
\hat{G}_k = g^{-1}\tilde{G}_k
\end{equation}
is an estimate of the gate set, up to a similarity transformation by the unobservable (see next section) matrix $B$.

It can happen that the SPAM gates that are implemented experimentally are not linearly independent within the bounds of sampling error. Then $\{|\rho_j\rb\}$ (and $\{\lb E_i|\}$) will not form a basis, and $g$ will have small or zero eigenvalues -- it will not be invertible. If this happens, the experimentalist must adjust the SPAM gates until $g$ is invertible. We will discuss this further in the next chapter.

\subsection{Gauge freedom}\label{gauge-freedom-section}
We now show that the matrix $B$ in Eq.~(\ref{g-inv-G}) is unobservable, as mentioned above. Recall that we assume no knowledge of $\mathcal{G}$, which includes $\lb E|$ and $|\rho\rb$. All experiments we consider are of the form $\lb E | G_{i_1} G_{i_2} ... G_{i_L}|\rho\rb$. (This includes measurements of $p_{ijk}$, since the $F_i$ are composed of elements of $\mathcal{G}$.) Transforming $\lb E | \rightarrow \lb E' | = \lb E| B$, $|\rho\rb \rightarrow |\rho'\rb = B^{-1}|\rho\rb$, $G_k \rightarrow G'_k = B^{-1}G_kB$, we find for a general expectation value: $\lb E' | G'_{i_1} G'_{i_2} ... G'_{i_L}|\rho'\rb = \lb E | G_{i_1} G_{i_2} ... G_{i_L}|\rho\rb$. Therefore, gates estimated by GST have a {\em gauge freedom} -- similarity transformation by a matrix, $B$. Any two sets of gates related to each other by a gauge transformation will describe a given set of measurements equally well (given that the states and measurements are transformed in the same way). They will be the same distance away from the actual gates (according to any distance metric) and, if they are physical, will have the same fidelity relative to the actual gate set.

Now consider the vectors
\begin{eqnarray}
|\tilde{\rho}\rb &\equiv& A|\rho\rb = \sum_i |i\rb\lb E|F_i|\rho\rb, \label{rhotilde}\\
\lb \tilde{E}| &\equiv& \lb E | B = \sum_j \lb E| F_j |\rho\rb \lb j|. \label{Etilde}
\end{eqnarray}
These vectors are componentwise identical, and the components are measurable quantities: $\lb i |\tilde{\rho}\rb = \lb \tilde{E} | i\rb = \lb E|F_i|\rho\rb$. They provide a way to construct a gate set consistent with our measurements, up to the gauge freedom $B$. Let
\begin{eqnarray}
|\hat{\rho}\rb &\equiv& g^{-1}|\tilde{\rho}\rb = B^{-1}|\rho\rb, \label{rhohat}\\
\lb\hat{E}| &\equiv& \lb\tilde{E}| = \lb E|B.\label{mhat}
\end{eqnarray}
As we saw before,
\begin{equation}
\hat{G}_k \equiv g^{-1}\tilde{G}_k =  B^{-1}G_kB.\label{Ghat}
\end{equation}

The gate set $\hat{\mathcal{G}} = \{|\hat{\rho}\rb,\lb\hat{E}|,\{\hat{G}_k\}\}$ consists entirely of measurable quantities: $\lb E | F_i G_k F_j |\rho\rb$ for $g$ and $\tilde{G}_k$, $\lb E| F_i |\rho\rb$ for $|\tilde{\rho}\rb$ and $\lb\tilde{E}|$. We can therefore construct it from experimental data. By gauge freedom, measurements in $\hat{\mathcal{G}}$ are equivalent to those in $\mathcal{G}$:
\begin{equation}
\lb\hat{E}|\prod_k \hat{G}_{i_k}|\hat{\rho}\rb = \lb E|B\left(\prod_k B^{-1}G_{i_k}B\right)B^{-1}|\rho\rb = \lb E|\prod_k G_{i_k}|\rho\rb.
\end{equation}
Hence, $\hat{\mathcal{G}}$ is indistinguishable from the true gate set $\mathcal{G}$.

However, since $B$ is never the identity, $\hat{\mathcal{G}}$ is not equal to $\mathcal{G}$. Since $B$ cannot be measured, the best we can do is find an estimate of $B$ that brings us as close as possible to some ``target'' gate set. The target may be chosen arbitrarily in accordance with the gauge freedom. Choosing it to be the {\em intended} experimental gate set allows us to compare the actual gates to ideal ones.

\subsection{Gauge optimization}\label{gauge-opt-section}

As we just saw, the gauge matrix $B_{ij} = \lb i| F_j |\rho\rb = \Tr\{P_i F_j(\rho)\}$ cannot be the identity. (This would require $F_j(\rho) = P_j \forall j$, which is impossible for a physical state.) Therefore, the gate set $\hat{\mathcal{G}}$ estimated by LGST is necessarily different from the true gate set $\mathcal{G}$. Unfortunately, experiments have no access to the gauge matrix $B$. But, since experiments look the same regardless of the gauge, we are free to choose a gauge that suits our purposes. This choice makes no practical difference -- a quantum computation in any gauge is indistinguishable from the same computation in another gauge.

Experimental qubits these days are quite good -- randomized benchmarking gate fidelities in excess of 99\% are routinely reported~\cite{barends_superconducting_2014}, In this case we know a priori that the measured gates will differ from an ideal (or target) set of gates by some very small error. A reasonable protocol then is to select the gauge such that the estimated gate set is as close as possible to this target gate set. The remaining differences are attributed to systematic gate error and sampling error.

To define closeness of quantum gates, we choose a suitable matrix norm on superoperators. For our purposes, the trace norm is sufficient~\cite{RBK2013}. Given a target set of gates $\mathcal{T} = \{|\tau\rb,\lb \mu |, \{T_k\}\}$ and our LGST estimate $\hat{\mathcal{G}} = \{|\hat{\rho}\rb,\lb\hat{E}|,\{\hat{G}_k\}\}$, we find the matrix $\hat{B}^*$ that minimizes the RMS discrepancy (trace norm):
\begin{equation}
\hat{B}^* = \argmin_{\hat{B}} \sum_{k=1}^{K+1} \Tr\left\{\left(\hat{G}_k - \hat{B}^{-1} T_k \hat{B}\right)^T \left(\hat{G}_k - \hat{B}^{-1} T_k \hat{B}\right)\right\}. \label{gaugeopt}
\end{equation}
The index $k$ runs over the entire gate set, and we also include the ``gate" $G_{K+1} \equiv |\rho\rb\lb E|$. This allows the states to be fitted as well.

Our final LGST estimate is then
\begin{eqnarray}
\hat{G}_k^* &=& \hat{B}^* \hat{G}_k (\hat{B}^*)^{-1}, \label{lgst-final1}\\
|\hat{\rho}^*\rb &=& \hat{B}^* |\hat{\rho}\rb, \label{lgst-final2}\\
\lb\hat{E}^*| &=& \lb\hat{E}|(\hat{B}^*)^{-1}. \label{lgst-final3}
\end{eqnarray}
This is the closest gate set to the target that is consistent with experiments as well as the allowable gauge freedom.

\subsection{Discussion}\label{discussion-section}
A few points are worth mentioning. First, we note that $\det(\hat{B})$ can be fixed by a requirement on the normalization of $|\hat{\rho}^*\rb$. We do not do this, but rather let $\hat{B}$ vary over the entire group of real, invertible matrices, $GL(d^2,\mathcal{R})$, letting the optimizer find the (not necessarily physical) gate set consistent with the data. In addition, Eq. (\ref{gaugeopt}) contains a nonlinear objective function ($B$ and $B^{-1}$ each appear twice), and therefore has multiple minima. As a result, the gauge optimization problem requires a starting point to be specified. This may lead to some unexpected consequences, which we now describe. 

Besides having multiple minima, the global minimum of the objective function, Eq. (\ref{gaugeopt}) is not unique. This indeterminacy is generic in quantum tomography, and appears in QST and QPT as well. In particular, we cannot distinguish $\lb U^\dag(E)| G |\rho\rb$ from $\lb E| G |U(\rho)\rb$ for any gate $G$, where $U$ is an arbitrary operation that commutes with $G$. (Depolarizing noise, for example.) The two sets $\{\lb U^\dag(E)| , |\rho\rb\}$, $\{\lb E|, |U(\rho)\rb\}$ are generally different. This indeterminacy can be recast as a type of gauge freedom, but in this case fixing the gauge provides no additional information about the true state and measurement. This has consequences for our analysis: an initialization error $|\mathcal{E}(\rho)\rb$ (e.g., a hot qubit) cannot be distinguished from a faulty measurement $\lb \mathcal{E}(E)|$ (e.g., dark counts). 

Typically, the starting point for numerical optimization of Eq. (\ref{gaugeopt}) is taken to be the target gauge matrix, defined as $\hat{B}^{(0)}_{ij} = \lb i| S_j |\tau\rb$. ($S_j$ is the target for $F_j$, composed of gates in $\mathcal{T}$ in the same way that $F_j$ is composed of gates in $\mathcal{G}$.) This starting point depends on the target $|\tau\rb$ for $|\rho\rb$ but does not depend on the target $\lb \mu|$ for $\lb E|$. As a result, gauge optimization will always produce $|\hat{\rho}^*\rb \approx |\tau\rb$, and will attribute any initialization error to error in $\lb \hat{E}^*|$, regardless of whether the error was actually in $\lb E|$ or $|\rho\rb$. This must be kept in mind when interpreting the results of GST for estimating states and measurements. The gate estimates themselves are not affected by this particular gauge freedom.

\section{Maximum likelihood estimation}\label{mle-section}

Linear inversion typically does not produce estimates that are physical (it is not constrained to do so). In contrast to QPT (see Sec.~(\ref{qpt-section}), Eq.~(\ref{qpt-ols-estimate})), it is also incapable of working with overcomplete data, which could be used to improve the estimate. MLE solves both of these problems. (Another approach is to find the closest physical gate set to the LGST estimate, but this is not optimal~\cite{RBK2013} and also cannot be extended to overcomplete data.) Since the objective function for MLE is nonlinear, the linear inversion estimate is still useful as a starting point for the optimization algorithm.\footnote{Standardized optimization methods exist for convex (linear, quadratic) objective functions, which have a single minimum~\cite{BoydVandenberghe}. When the objective is nonlinear, with many local minima, there are no existing techniques that are guaranteed to find the global minimum. We must therefore use local optimization techniques and rely on a good starting point to put us close to the global minimum.}

Our goal is to estimate the true probabilities $p_{ijk}$ in Eq. (\ref{pijk-0}) based on a set of measurements $m_{ijk}$, subject to physicality constraints. MLE is the natural way to do this: the best estimate is found by fitting to experimental data a theoretical model of the probability of obtaining that data. The resulting estimates $\hat{p}_{ijk}$ are used to find the most likely set of gates $\hat{\mathcal{G}}$ that produced the data.

We begin by parameterizing the $4 \times 4$ estimate matrices $\hat{G}$, as well as the state and measurement estimates $\hat{\rho}$, $\hat{E}$, in terms of a vector of parameters, $\vec{t}$: $\hat{G}(\vec{t})$, $|\hat{\rho}(\vec{t})\rb$, $\hat{E}(\vec{t})$. (Although we have written the same vector $\vec{t}$ as the argument in each matrix, the different gates and states depend on non-overlapping subsets of parameters in $\vec{t}$.) There are several possibilities for the parameterization, which we discuss in detail later. The important point for now is that the parameterization is either linear or quadratic in $\vec{t}$: each matrix element of $G(\vec{t})$, $|\rho(\vec{t})\rb$, $E(\vec{t})$ is a polynomial of order 1 or 2 in the parameters $\{t_i\}$. There are $d^4=16$ parameters for each gate $G_k$, $k=1,\ldots,K$ ($k=0$ is not parameterized), and $d^2=4$ parameters each for $E$ and $\rho$.

A number of constraints reduce the total number of independent parameters, as we discuss later. The minimal case of GST on one qubit requires a gate set $\mathcal{G}$ with $K=2$. This gives $K d^4 + 2 d^2 =  40$ parameters. Equality constraints (trace preservation requirement) reduce the number of parameters by 4 per gate, and by 1 for $\rho$. Thus we are left with $31$ free parameters in the minimal instance of GST.

Putting everything together, the estimates $\hat{p}_{ijk}$ are written in terms of the parameter vector $\vec{t}$, as
\begin{equation}
\hat{p}_{ijk}(\vec{t}) =  \lb \hat{E}(\vec{t})|\hat{F}_i(\vec{t})\hat{G}_k(\vec{t})\hat{F}_j(\vec{t}) |\hat{\rho}(\vec{t})\rb. \label{phat}
\end{equation}
Depending on the parameterization choice (discussed below), each of the gates and states in Eq.~(\ref{phat}) is either a linear or quadratic function of its parameters. The estimator $\hat{p}_{ijk}$ is therefore a homogeneous function of order 5 (linear parameterization) or 10 (quadratic parameterization).

MLE proceeds by finding the set of parameters $\vec{t}$ that minimizes an objective function. The objective, or likelihood function, is the probability distribution we assume produced the data. The most general likelihood function for the experiment we have described above is~\cite{RBK2013}
\begin{equation}
L(\hat{\mathcal{G}}) = \prod_{ijk} (\hat{p}_{ijk})^{m_{ijk}}(1-\hat{p}_{ijk})^{1-m_{ijk}}.
\end{equation}
Although several workers~\cite{RBK2013,Yudan} have successfully used this likelihood function with their experimental data, it is usually more convenient to use a simpler form. To this end, we invoke the central limit theorem to rewrite the likelihood as a normal distribution,
\begin{equation}
L(\hat{\mathcal{G}}) = \prod_{ijk}\exp\left[-\left(m_{ijk}-\hat{p}_{ijk}\right)^2/\sigma_{ijk}^2\right],
\end{equation}
where $\sigma^2 = p(1-p)/n$ is the sampling variance in the measurement $m$.\footnote{When performing numerical optimization, we typically approximate $p \approx m$ rather than $p \approx \hat{p}$ in the sampling variance, writing $\sigma^2 = m(1-m)/n$. This keeps the objective function from blowing up.}

Because the logarithm function is monotonic, maximizing the likelihood $L$ is equivalent to minimizing the negative log-likelihood $l = -\log L$. With a normal likelihood function, the problem reduces to weighted least-squares:
\begin{equation}
\mathrm{Minimize}:\;\;\; l(\hat{\mathcal{G}}) = \sum_{ijk}\left(m_{ijk}-\hat{p}_{ijk}(\vec{t})\right)^2/\sigma_{ijk}^2.
\end{equation}
Or, rewriting the estimators $\hat{p}_{ijk}$ in terms of the gate estimates, Eq. (\ref{phat}), the problem we need to solve is minimization over the parameters $\vec{t}$ of
\begin{equation}
 l(\hat{\mathcal{G}}) = \sum_{ijk}\left(m_{ijk}-\lb \hat{E}(\vec{t})|\hat{F}_i(\vec{t})\hat{G}_k(\vec{t})\hat{F}_j(\vec{t}) |\hat{\rho}(\vec{t})\rb\right)^2/\sigma_{ijk}^2. \label{wlsq}
\end{equation}
As we saw above, the estimator $\hat{p}_{ijk}$ is of order 5 or 10 in the parameters. Therefore this objective function is 10-th to 20-th order in the parameters, making this a non-convex optimization problem.

It remains to express $\hat{p}_{ijk}$ in terms of the parameters $\vec{t}$, so that we have an explicit form for $ l(\hat{\mathcal{G}})$ in Eq. (\ref{wlsq}) in terms of $\vec{t}$. There are two parameterizations for gates that are commonly used: (1) The Pauli Process ($\chi$) Matrix representation and (2) the Pauli Transfer Matrix (PTM) representation. Each has its advantages, so we present both.

\subsection{Process matrix optimization problem}\label{process-matrix-optimization-section}

The process matrix $\chi_G$ for a gate $G$ is defined in terms of the gate's action on an arbitrary state $\rho$ to produce a new state $G(\rho)$ as in Section~\ref{process-matrix-section},
\begin{equation}
G(\rho) = \sum_{i,j=1}^{d^2} (\chi_G)_{ij} P_i \rho P_j.
\end{equation}
The gate action is expressed in terms of Pauli operators (see Section ~\ref{quantops}) $P_i$, $P_j$ acting on the state.

The $\chi$ matrix must be Hermitian positive semidefinite. This requirement follows from the Hermiticity of density matrices, $\rho^\dag = \rho$, $G(\rho)^\dag = G(\rho)$, and from the requirement of positive probabilities: $p_i = \Tr\{|i\rangle\langle i| \rho\} \geq 0$ for any pure state $|i\rangle$ and any density matrix $\rho$ (see Sec.~(\ref{physicality-section})). Any Hermitian positive semidefinite matrix can be written in terms of a Cholesky decomposition, $\chi = T^\dag T$, where $T$ is a lower-diagonal complex matrix with reals on the diagonal:
\begin{eqnarray}
T &=& \left(\begin{array}{cccc} t_1  &          0   &           0      &       0\\
    t_5+i t_6 &  t_2   &        0    &         0\\
    t_{11}+i t_{12} & t_7+i t_8  & t_3     &     0\\
    t_{15}+i t_{16} &t_{13}+i t_{14} & t_9+i t_{10} &t_4\end{array}\right). \label{T}
\end{eqnarray}
In addition, there will be a constraint on $\chi$ due to the requirement of trace preservation of the state after the application of the gate: $\Tr\{G(\rho)\} = \Tr\{\rho\}$. This leads to the completeness condition, $\sum_{ij}\chi_{ij}P_jP_i = I$, which is equivalent to the following four equations.
\begin{equation}
\sum_{ij}\chi_{ij}\Tr\{P_iP_kP_j\} = \delta_{k0},\;\;\; k=1,\ldots,4.
\end{equation}
This constraint reduces the number of free parameters for $\chi$ from 16 to 12. In practice, we leave the 16 parameters and introduce the constraint as an equality constraint in the numerical optimizer.

We now discuss parameterization of $E$ and $\rho$. Both are Hermitian positive semidefinite, so we may parameterize them via Cholesky decomposition in the same way as $\chi$. (Note only that they are $2 \times 2$ rather than $4 \times 4$ matrices, so that each will have 4 parameters.) The state $\rho$ has unit trace, and the measurement $E$ must be such that $I-E$ is also positive semidefinite. These conditions introduce additional constraints for optimization.

We can now write down an explicit form for the weighted least-squares objective function, Eq. (\ref{wlsq}), in terms of the parameter vector, $\vec{t}$. We expand the estimator $\hat{p}_{ijk}$ in terms of the $\chi$ matrices for its constituent gates using Eq. (\ref{chi}),
\begin{eqnarray}
\hat{p}_{ijk} &=& \lb \hat{E}(\vec{t})|\hat{F}_i(\vec{t})\hat{G}_k(\vec{t})\hat{F}_j(\vec{t}) |\hat{\rho}(\vec{t})\rb \nonumber \\
&=& \Tr\{\hat{E} \hat{F}_i(\hat{G}_k(\hat{F}_j(\rho)))\} \nonumber \\
&=& \sum_{mnrstu} (\chi_{F_i})_{tu}(\chi_{G_k})_{rs}(\chi_{F_j})_{mn} \Tr\{E P_t P_r P_m \rho P_n P_s P_u\}. \label{pchi}
\end{eqnarray}
Remember that each $\chi$ matrix as well as $\rho$ and $E$ is written as $T^\dag T$ in terms of its own set of parameters (the $\chi$s are $4 \times 4$ and $\rho$ and $E$ are $2\times 2$), and therefore $\hat{p}_{ijk}$ is generally a homogeneous function of order 10 in these parameters (it consists of 3 $\chi$-matrices plus $\rho$ and $E$). Also, it may be the case that some of the $F_i, F_j$ are composed of more than one gate $G \in \mathcal{G}$. In this case $\chi_{F_i}$ and $\chi_{F_j}$ can be decomposed further into process matrices for the gates in $F_i, F_j$, with corresponding additional Pauli matrix terms appearing inside the trace. This makes $\hat{p}_{ijk}$ a correspondingly higher order polynomial. For this reason we would like to avoid defining SPAM gates as combinations of gates in the gate set.

\subsubsection{Problem statement}
We can now state the optimization problem as follows.
\begin{eqnarray}
\mathrm{Minimize:}&&\;\;\; \sum_{ijk}\left(m_{ijk}-\sum_{mnrstu} (\chi_{F_i})_{tu}(\chi_{G_k})_{rs}(\chi_{F_j})_{mn} \Tr\{E P_t P_r P_m \rho P_n P_s P_u\}\right)^2/\sigma_{ijk}^2 \nonumber \\
\mathrm{Subject\;to:}&&\;\;\; \sum_{mn}(\chi_G)_{mn}\Tr\{P_m P_r P_n\} - \delta_{0r} = 0, \;\;\; \forall\; G \in \mathcal{G} \nonumber \\
&&\;\;\; \Tr\{\rho\} = 1, \nonumber \\
&&\;\;\; I - E \succcurlyeq 0 \nonumber
\end{eqnarray}

The wavy inequality denotes positive semidefiniteness. For each $G\in\mathcal{G}$ there is a $4\times 4$ matrix $\chi_G = T_G^\dag T_G$, where $T_G$ is a lower diagonal complex matrix with real entries on the diagonal, Eq. (\ref{T}). There are 16 free parameters $t_i$ for each $G \in \mathcal{G}$, and each matrix element of $\chi_G$ is a 2nd order polynomial in the $\{t_i\}$.

The matrices $\chi_F$ must be expressed in terms of matrices $\chi_G$. In the second example in Sec.~\ref{example} above, this is trivial since $\mathcal{F} = \mathcal{G}$. In the first example the only non-trivial case is $F_4 = G_1\circ G_1$. The simplest way to find $\chi_{F_4}$ is to calculate the PTM for $F_4$ by multiplying the PTMs for $G_1$, and then transforming back to the $\chi$ representation. This can be implemented with a simple numerical routine.

$E$ and $\rho$ are parameterized similarly to $\chi$ as $T^\dag T$, only they are $2 \times 2$ matrices. Each therefore contains 4 parameters.

Possible starting points are the target gate set or the LGST estimate (rather, the closest physical gate set to the LGST estimate). Typically LGST provides a better starting point than the target gate set.

\subsection{Pauli transfer matrix optimization problem}\label{ptm-optimization-section}

Next we discuss parameterization in terms of Pauli Transfer Matrices (PTMs). As we saw in Ch.~\ref{math-chapter}, the composition of gates is represented as matrix multiplication of PTMs. This avoids the cumbersome trace terms appearing in equations such as Eq. (\ref{pchi}). In addition, the PTM for each gate may be parameterized linearly rather than quadratically as we did for $\chi$. This reduces the order of the objective function by a factor of 2. The drawback is that the positivity constraint -- corresponding to positive-definiteness of $\chi$ that we imposed above via Cholesky decomposition -- has a more complicated structure in terms of PTMs. It is expressed by the requirement of positive semidefiniteness of the so-called Choi-Jamiolkowski (CJ) matrix representing the quantum map, see Sec.~\ref{ptm-physicality}. Imposing this type of constraint without Cholesky decomposition is difficult to do with standard nonlinear optimization techniques, but can be done using semidefinite programming (SDP)~\cite{Chow2012SI} if the objective function can be made quadratic.

The Pauli Transfer Matrix $R_G$ for a gate $G$ is defined in terms of the gate's action on Pauli matrices (see Section~\ref{ptm-section}),
\begin{equation}
(R_G)_{ij} = \Tr\{P_i G(P_j)\}. \label{PTM-R}
\end{equation}
The PTM contains the same information about the map as does $\chi$. It is a $d^2 \times d^2$ real-valued matrix with elements restricted to the interval $R_{ij}\in[-1,1]$, and is generally not symmetric. Like $\chi$, the PTM has 16 parameters. The trace-preserving condition is $R_{0j} = \delta_{0j}$. These 4 equations reduce the number of parameters to 12.

In the PTM representation, density matrices $\rho$ in Hilbert space are written as vectors in Hilbert-Schmidt space and denoted as $|\rho\rb$. The matrix elements (entries) of a general vector $|\rho\rb$ are defined as
\begin{equation}
\rho_i = \lb i | \rho\rb = \Tr\{P_i \rho\}. \label{rhovec}
\end{equation}
The left-hand term in the equality is the $i$-th component of the $4 \times 1$ vector $|\rho\rb$. The middle term is this same component written in Dirac notation. In the right-hand term, $\rho$ is the density matrix in standard $2 \times 2$ representation, and $P_i$ is a Pauli matrix.

Using these definitions, the state $G(\rho)$ after application of the gate $G$ can be written as a vector $|G(\rho)\rb$ in Hilbert-Schmidt space resulting from matrix multiplication by $R_G$:
\begin{equation}
|G(\rho)\rb = R_G|\rho\rb
\end{equation}
The standard density matrix form of $G(\rho)$ can be recovered from the vector $|G(\rho)\rb$ using the definition of matrix elements, Eq. (\ref{rhovec}),
\begin{equation}
\Tr\{P_i G(\rho)\} = \lb i | G(\rho)\rb,
\end{equation}
and the expansion of any density matrix in terms of Pauli matrices,
\begin{equation}
G(\rho) = \sum_{i=1}^4 \Tr\{P_i G(\rho)\} P_i.
\end{equation}

In Eq. (\ref{wlsq}), the weighted least-squares objective function has already been written in terms of PTMs. This was implicit in the superoperator notation we used to derive that equation. In terms of the $R$-matrix notation for PTMs, we rewrite Eq. (\ref{wlsq}) as
\begin{eqnarray}
 l(\hat{\mathcal{G}}) = \sum_{ijk}\left(m_{ijk}-\lb \hat{E}(\vec{t})|\hat{R}_{F_i}(\vec{t})\hat{R}_{G_k}(\vec{t})\hat{R}_{F_j}(\vec{t}) |\hat{\rho}(\vec{t})\rb\right)^2/\sigma_{ijk}^2. \label{wlsq-R}
\end{eqnarray}
We can parameterize each matrix $\hat{R}$ linearly in terms of $\vec{t}$. This means each $R_{ij} = t_s$ for some index $s$. The term in double-brackets, $\lb\ldots\rb$, in Eq. (\ref{wlsq-R}) is a scalar given by applying the indicated matrix multiplications of $\hat{R}$-matrices to the vector $|\hat{\rho}\rb$, and then scalar multiplying by $\lb\hat{E}|$. 

The state $|\hat{\rho}\rb$ and measurement $\lb \hat{E}|$ should also be parameterized linearly, and the constraints (positive semidefiniteness, hermiticity, trace-preservation of $\rho$) can be imposed during optimization.

Since each $\hat{R}$ is linear in the parameters, $\vec{t}$, the matrix product in Eq. (\ref{wlsq-R}) is 5th order in the parameters, $\vec{t}$. Therefore the objective function, Eq. (\ref{wlsq-R}) is a 10th-order polynomial in $\vec{t}$.

As mentioned earlier, the positivity constraint is expressed as the positive semidefiniteness of the Choi-Jamiolkowski matrix, Eq.~(\ref{choi1}).

\subsubsection{Problem statement}
\begin{eqnarray}
\mathrm{Minimize:}&&\;\;\; \sum_{ijk}\left(m_{ijk}-\lb \hat{E}(\vec{t})|\hat{R}_{F_i}(\vec{t})\hat{R}_{G_k}(\vec{t})\hat{R}_{F_j}(\vec{t}) |\hat{\rho}(\vec{t})\rb\right)^2/\sigma_{ijk}^2 \nonumber \\
\mathrm{Subject\;to:}&&\;\;\; \rho_G = \frac{1}{d^2}\sum_{i,j=1}^{d^2} (R_G)_{ij} P_j^T \otimes P_i \succcurlyeq 0, \;\;\; \forall\; G \in \mathcal{G} \nonumber \\
&&\;\;\; (R_G)_{0i} = \delta_{0i} \;\;\; \forall\; G \in \mathcal{G} \nonumber \\
&&\;\;\; (R_G)_{ij} \in [-1,1] \;\;\; \forall\;G,i,j \nonumber \\
&&\;\;\; \Tr\{\rho\} = 1, \nonumber \\
&&\;\;\; I - E \succcurlyeq 0 \nonumber
\end{eqnarray}
For each $G\in\mathcal{G}$ the $4\times 4$ matrix $R_G$ is parameterized linearly, $(R_G)_{ij} = t_s$ for some index $s$. $|\rho\rb$ is a $4\times 1$ vector and $\lb E|$ is a $1\times 4$ vector. $\rho$ and $E$ are parameterized linearly in such a way that they are hermitian and positive semidefinite. The components $\rho_i$ of $|\rho\rb$ are given as $\rho_i = \Tr\{P_i \rho\}$ and similarly for $\lb E|$.

The product in double brackets in the objective function is a scalar calculated by matrix multiplication of the Pauli transfer matrices for the gates as indicated. For the first example in Sec.~\ref{example}, the matrix $R_{F_4} = (R_{G_1})^2$. Note that this choice increases the order of the objective function in the parameters, $\vec{t}$, since $G_1$ appears twice in the same $F$-gate. As a result, the 2nd example gate set in Sec.~\ref{example} is a better choice.

This version of the optimization problem has the advantage over the $\chi$-matrix version in that the objective function is a 10th-order rather than a 20th-order polynomial in $\vec{t}$. This is still a highly nonlinear objective function. The tradeoff is the requirement of positive-semidefiniteness of the matrix $\rho_G$ for each $G$, which is not necessary in the $\chi$ matrix approach. This requirement naturally suggests a solution in terms of a semidefinite program (SDP). In Ref.~\cite{Merkel2013}, this is achieved by replacing the objective function by a quadratic approximation in order to recast the problem as a convex optimization problem, which is then solved by SDP.  (Recall that convex optimization requires the objective function to be of order 2 or less in the optimization parameters.) This approximation amounts to linearizing the triple-gate-product in the estimator $\hat{p}_{ijk}$ about the target gate set, and eliminating the $t$-dependence of $\rho$ and $E$. The latter can be done either assuming the state and POVM are perfect, or by introducing imperfect ones by hand or from the LGST estimate. 

Finally, we note that a different gate set, such as the linear inversion estimate, may be used as the starting point for nonlinear optimization, or as the point about which the gate set is linearized for convex optimization.

\chapter{Implementing GST}\label{implementation-chapter}

The data-processing challenge of GST boils down to that of solving a nonlinear optimization problem. We are presented with measurements $m_i$ of a set of probabilities $p_i = \lb E | G_i(\rho)\rb$, $i=0,...,K$. (Using a simplified notation with a single index $i$.) Based on these measurements alone, we would like to find the gate set $\mathcal{G}=\{|\rho\rb,\lb E|,\{G_i\}\}$ that produced the data. Practically speaking, we would like to find the best estimate $\hat{p}_i$ of the true probabilites $p_i$, given the experimentally measured values $m_i$, where the $\hat{p}_i$ are functions of the gate set (i.e., of the parameters used to define the gate set). This estimate should correspond to a physical (CPTP -- see Sec.~\ref{physicality-section}) set of gates.

Typically, each of the $K$ experiments is repeated a sufficiently large number of times $N$ that the central limit theorem applies. Therefore $m_i$ can be considered a sample from a Gaussian distribution.\footnote{More accurately, $m_i$ is a sample from a binomial distribution, which can be approximated as Gaussian when $p_i N$ and $(1-p_i)N$ are not too small. In a hypothetical  ideal experiment with no intrinsic SPAM errors, it can happen that $p_i \approx 0$ or $1$ for some values of $i$ when the gate error is low. The Gaussian approximation then breaks down and we must use the full binomial objective function. Since real experiments are not perfect (noise affects initialization, readout, etc.) this should not be an issue.} Then, as we have seen, the problem of estimating $\hat{p}_i$ can be rephrased as maximum-likelihood estimation (MLE) with an objective function (log-likelihood) that is quadratic in the $\hat{p}_i$. Furthermore, if $G_i$ is a single gate (this is the case in QPT), the objective function is quadratic in the gate parameters, and thus convex. (A convex function has only a single local minimum, the global minimum, in its domain of definition.) In this case, the problem can be solved using standard convex optimization techniques~\cite{Chow2012SI,BoydVandenberghe}.

Computation of the QPT estimate is therefore a solved problem (aside from some technicalities, see Ref.~\cite{RBK2010}). For a linearly-parameterized gate set, it is straightforward (though possibly computationally intensive) to determine the most likely CPTP gates that produced the data, as well as the errors in the estimate. Unfortunately, QPT does not solve the correct problem. For self-consistency, we require state preparation and measurement (SPAM) gates to be included in the $\{G_i\}$, making the gate set at least 3rd order in the gate parameters. As a result the objective function is no longer convex (it has many local minima) and the problem is no longer solvable by standard convex optimization techniques. Instead one has to choose a combination of approximate and iterative methods that one hopes can locate the global optimum.

Our goal in this chapter is to illustrate the effectiveness of GST. We use the simplest optimization methods that get the job done. Thus, we implement full nonlinear optimization in Matlab, choosing reasonable settings for the optimization routines but making no effort to optimize these settings. We find this is sufficient to illustrate the main features of GST.

We begin by summarizing the steps required to implement GST, including how to gather and organize the data, and some tips on the analysis. We then present the results of GST for a single qubit using simulated data with simplified but realistic errors. We compare the performance of maximum likelihood GST (ML-GST) to that of QPT for varying levels of coherent error, incoherent error, and sampling noise. We corroborate the result of Ref.~\cite{Merkel2013} that coherent errors are poorly estimated by QPT near QEC thresholds while GST is accurate in this regime.

In Chapter~\ref{derivation-chapter}, we described two versions of ML-GST, a nonlinear optimization problem based on the process matrix (Sec.~\ref{process-matrix-optimization-section}) and a semidefinite program (SDP) based on the Pauli transfer matrix (Sec.~\ref{ptm-optimization-section}). The numerical results in the present chapter were obtained using the process matrix - based approach. Due to the high nonlinearity of the objective function in this approach, the LGST estimate was essential as a starting point for MLE and provided better estimates than using the target gate set as a starting point.

\section{Experimental protocol} \label{experimental-protocol}
Recalling the definitions from the last chapter, $\mathcal{G} = \{G_0,G_1,\ldots,G_K\}$ is the gate set, with $G_0$ denoting the ``null'' gate - do nothing for no time - a perfect identity. $\mathcal{F} = \{F_1,\ldots,F_N\}$ is the SPAM gate set. It is used to construct a complete basis of starting states and measurements from a given particular starting state $\rho$ and measurement operator $E$. Each gate $F_i \in \mathcal{F}$ is composed of gates in $\mathcal{G}$. For a single qubit, $d=2$, we must have $N\geq 4$. For simplicity, we can take $N=4$.

The experimental protocol is as follows.

\begin{enumerate}
\item[1.] Initialize the qubit to a particular state $|\rho\rb$. In most systems, the natural choice for $\rho$ is the ground state of the qubit, $\rho = |0\rangle\langle 0|$.
\item[2.] For a particular choice of $i,j \in \{1,\ldots,N\}$, $k\in\{0,\ldots,K\}$, apply the gate sequence $F_i \circ G_k \circ F_j$ to the qubit. Remember that the $F$ gates are composed of $G$s. So the gate sequence applied in this step is a sequence of gates $G \in \mathcal{G}$.
\item[3.] Measure the POVM $E$. $E$ is required to be a positive semidefinite Hermitian operator, such that $I-E$ is also positive semidefinite. ($I$ is the identity.) As is the case for $\rho$, the natural choice for $E$ in most systems is $E = |0\rangle\langle 0|$. Sometimes $E=|1\rangle\langle 1|$ is used. 
\item[4.] Repeat steps 1-3 a large number of times, $n$. Typically, $n = $1,000 - 10,000. For the $r$-th repetition, record $n_r = 1$ if the measurement is success (i.e., the measured state is $|0\rangle\langle 0|$), and $n_r = 0$ if the measurement is failure (i.e., the measured state is not $|0\rangle\langle 0|$).
\item[5.] Average the results of step 4. The result, $m_{ijk} = \sum_{r=1}^n n_r/n$, is a measurement of the expectation value $p_{ijk} = \lb E|F_i G_k F_j |\rho\rb$. It is a random variable with mean $p_{ijk}$ and variance $p_{ijk} (1-p_{ijk}) / n$.
\item[6.] Repeat steps 1-5 for all $i,j \in \{1,\ldots,N\}$, $k\in\{0,\ldots,K\}$.
\item[7.] {\em Optional --} Repeat steps 1-5 to measure the expectation values $p_i = \lb E| F_i |\rho\rb$. Since typically $F_0 = G_0 = \{\}$, this data will already be contained in the measurements of $p_{ijk}$. However, it helps to have an independent measurement if possible.
\end{enumerate}

\section{Organizing and verifying the data}

The procedure above gives measurements of the following quantities.
\begin{eqnarray}
&\lb E| F_i \circ G_k \circ F_j|\rho\rb,&\nonumber \\
&\lb E| F_i \circ F_j|\rho\rb,&\nonumber \\
&\lb E| F_i |\rho\rb,&\nonumber
\end{eqnarray}
for $k=1,\ldots,K$ and $i,j=1,\ldots,N$.

The experimental data should then be organized into matrices, as follows (see Eqs.~(\ref{pijk}), (\ref{gram}), (\ref{rhotilde})-(\ref{Etilde})).
\begin{eqnarray}
(\tilde{G}_k)_{ij} & = & \lb E| F_i G_k F_j|\rho\rb, \label{exp2}\\
g_{ij} & = & \lb E| F_i F_j|\rho\rb, \label{exp1}\\
|\tilde{\rho}\rb_i & = & \lb E| F_i |\rho\rb = \lb \tilde{E}|_i, \label{exp3}
\end{eqnarray}
where in an abuse of notation we have written $\lb E|\ldots|\rho\rb$ to stand for the {\em measured} values of these quantities rather than the true values.

Once data is obtained, and before proceeding further, we must check that the Gram matrix, $g_{ij} = \lb E |F_i F_j|\rho\rb$, is nonsingular, so that it may be inverted to find the estimate as in Eqs. (\ref{rhohat})-(\ref{Ghat}). We want the smallest magnitude eigenvalue to be as large as possible. This is because the sampling error on the estimate scales roughly as the inverse of the smallest eigenvalue of the Gram matrix (multiplied by the sampling error in the data). A good rule of thumb is that no eigenvalue be less than 0.1 in absolute value. (Then for $N=2000$ samples the sampling error in the estimate is bounded at $5\%$.) 

If the eigenvalues of the Gram matrix are very small, this indicates that the SPAM gates are only marginally linearly independent. Viewed as vectors, they are highly overlapping. In this case, the experimenter must go back and tweak the knobs of the experiment to make the SPAM gates more orthogonal. Then the Gram matrix must be measured again and its eigenvalues checked. This process is repeated until a suitable Gram matrix is obtained.\footnote{Typically experimentalists have other means at their disposal to ensure orthogonal, or nearly orthogonal, gate rotation axes. Thus it is not difficult to obtain an invertible Gram matrix in practice.}

\section{Linear inversion}

Once we have checked that the experimental Gram matrix is invertible, we apply its inverse to the data matrices in Eqs.~(\ref{exp2})-(\ref{exp3}), following the procedure in Section~\ref{derivation-lgst}. (Note that $g^{-1}$ is applied to all data matrices except the vector $\lb \tilde{E}|$.) We obtain the following estimates for the gates and states.
\begin{eqnarray}
|\hat{\rho}\rb &=& g^{-1}|\tilde{\rho}\rb, \label{lgst-est1}\\
\lb\hat{E}| &=& \lb\tilde{E}|,\label{lgst-est2}\\
\hat{G}_k &=& g^{-1}\tilde{G}_k.\label{lgst-est3}
\end{eqnarray}
Since $g \equiv \tilde{G}_0$, the LGST estimate of the null gate is exactly the identity, as it should be.

The gate set estimated in this way, $\hat{\mathcal{G}} = \{|\hat{\rho}\rb,\lb\hat{E}|,\{\hat{G}_k\}\}$, is in a different gauge from the actual gate set (Sec.~\ref{gram-ab-section} - \ref{gauge-opt-section}). To compensate, we transform to a more useful gauge. Since we do not know the actual gate set, only the intended (target) one, the most useful gauge is the one that brings the estimated gate set as close as possible (based on some distance metric) to the target. The difference between the final estimate and the target gate set is then a measure of the error in the actual gate set (how far away it is from what was intended).

The gauge transformation is found by solving the optimization problem defined in Eq.~\ref{gaugeopt}. The resulting gauge matrix, $\hat{B}^*$, is then applied to Eqs.~(\ref{lgst-est1}) - (\ref{lgst-est3}). The final LGST estimate is given by Eqs.~(\ref{lgst-final1}) - (\ref{lgst-final3}), which we reproduce here:
\begin{eqnarray}
\hat{G}_k^* &=& \hat{B}^* \hat{G}_k (\hat{B}^*)^{-1}, \\
|\hat{\rho}^*\rb &=& \hat{B}^* |\hat{\rho}\rb, \\
\lb\hat{E}^*| &=& \lb\hat{E}|(\hat{B}^*)^{-1}.
\end{eqnarray}

The linear inversion protocol is fairly easy to implement numerically and is useful for providing quick diagnostics without resorting to more computationally intensive estimation via constrained optimization. Since the LGST estimate is not generally physical,\footnote{There is no natural way to put physicality constraints into the LGST protocol. One option is to find the closest physical gate set to the LGST estimate, but this is suboptimal to MLE.} the information obtained is somewhat qualitative. Nevertheless, large enough errors can be easily detected with this approach.

More importantly, LGST is useful as a starting point for MLE. Since the starting point must be physical, while LGST is not, we should use the closest physical gate set to the LGST estimate. Such a gate set can be found in a similar way to the gauge optimization procedure described above, choosing a metric such as the trace norm as a measure of distance between gates.

\section{Estimation results and analysis for simulated experimental data}

In this section we present results of maximum likelihood - based GST (ML-GST) for several examples using simulated data, and compare to ML-QPT. The simulated errors are of three different types, representing coherent and incoherent gate errors as well as intrinsic SPAM errors. The MLE approach was discussed in Section~\ref{mle-section}.

Maximum likelihood estimation provides a convenient way to handle physicality constraints as well as overcomplete data (we will get to that later). Although it has been argued that MLE is suboptimal to Bayesian methods~\cite{RBK2010}, it is still the method of choice for GST and QPT~\cite{Chow2009SI, Chow2012, Merkel2013, RBK2013}. We shall see that MLE is highly accurate.

As a model system, we consider a single qubit with available gates consisting of rotations about two orthogonal axes, which we label X and Y. This is the case in many existing qubit implementations~\cite{barends_superconducting_2014,Chow2012}. Furthermore, we assume the qubit can be prepared in its ground state, $\rho = |0\rangle\langle 0|$ and measured in the $Z$-basis, distinguishing $E=|1\rangle\langle 1|$ and $I-E=|0\rangle\langle 0|$. This state and measurement may be faulty, and we will study the effect on estimation due to these errors.

The results in this section were obtained using gate set 2 in Sec.~\ref{example},
\begin{eqnarray}
\mathcal{G} &=& \{\{\},X_{\pi/2},Y_{\pi/2},X_{\pi}\},\\
\mathcal{F} &=& \mathcal{G}.
\end{eqnarray}

To illustrate the effect of systematic errors (all errors with the exclusion of statistical sampling noise are systematic errors), we use depolarizing noise as an example of incoherent environmental noise, and over-rotations in the Y-gate as an example of a coherent control error. Depolarizing noise is defined by the map
\begin{equation}
G_{dep}(\rho) = (1-3p)\rho + p(X\rho X + Y\rho Y + Z\rho Z) \label{depol-gate}
\end{equation}
For a 50~$\mathrm{ns}$ gate, a depolarizing parameter $p = 0.005$ corresponds to a decoherence time of 2.5~$\mathrm{\mu s}$. In our numerical experiments we apply $G_{dep}$ after every gate where depolarizing noise is required.

Over-rotation errors are obtained by applying the map,
\begin{equation}
G_{rot}(\rho) = \exp\left(-i \frac{\epsilon}{2}\hat{n}\cdot \vec{\sigma}\right) \rho \exp\left( i \frac{\epsilon}{2}\hat{n}\cdot \vec{\sigma} \right),
\end{equation}
after every gate that should have the error, in our case the Y-gate. The parameter $\epsilon$ is the angle of over-rotation, $\hat{n}=\hat{y}$ is the rotation axis, and $\vec{\sigma} = (X,Y,Z)$ is a vector of Pauli matrices.

In addition to these systematic errors, there will be noise due to finite sampling statistics. As an example, $N = 2000$ samples per experiment produces a sampling error in the data that is upper-bounded by $1/(4\sqrt{N}) = 0.005$. 

\subsection{Systematic errors}\label{examples-subsection}

We first consider systematic errors only, no sampling error. This is useful for examining how well GST can do relative to QPT {\em in principle}. In practice, sampling error will contaminate both GST and QPT estimates, and efforts must be taken to reduce it. This can be done either by taking more samples (which can be impractical) or performing more independent experiments.

We calculate estimates using the process matrix formulation of the MLE problem, see Sec.~\ref{process-matrix-optimization-section}. This is a nonlinear optimization problem requiring a starting point to be specified. For the starting point, we use the LGST estimate for GST and the target gate set for QPT. More precisely, for GST we find the closest physical gate set to the LGST estimate (based on the trace norm) and use that for a starting point.

Since the number of samples is taken as infinite (corresponding to zero sampling error), we use a standard (non-weighted) least squares objective function, 
\begin{equation}
l(\hat{\mathcal{G}}) = \sum_{i=1}^{N_{exp}} \left(m_i - \hat{p}_i\right)^2,
\end{equation}
where $N_{exp}$ is the number of experiments (16 for QPT, 84 for GST) and $m_i$ are the measured values of the true probabilities $p_i$, of which $\hat{p}_i$ are our estimates. Since the sampling error is zero, $m_i = p_i$. 

The estimates in this section were generated using Matlab's built-in optimization function {\em fmincon}, running the active-set optimization algorithm. Optimization time varied between 2-4 minutes for each GST estimate on an Intel(R) Core(TM) i5-2500K (3.3GHz) processor. Optimizer settings were options.TolFun = $10^{-12}$, options.TolCon = $10^{-6}$, options.MaxFunEvals = 15000. Typically the objective function and constraint tolerances were satisfied before the maximum number of function evaluations was reached.

\subsubsection{Example 1: Over-rotation (coherent) error}

\begin{figure}
\includegraphics[width=16.4cm, trim = 2cm 0 0 0]{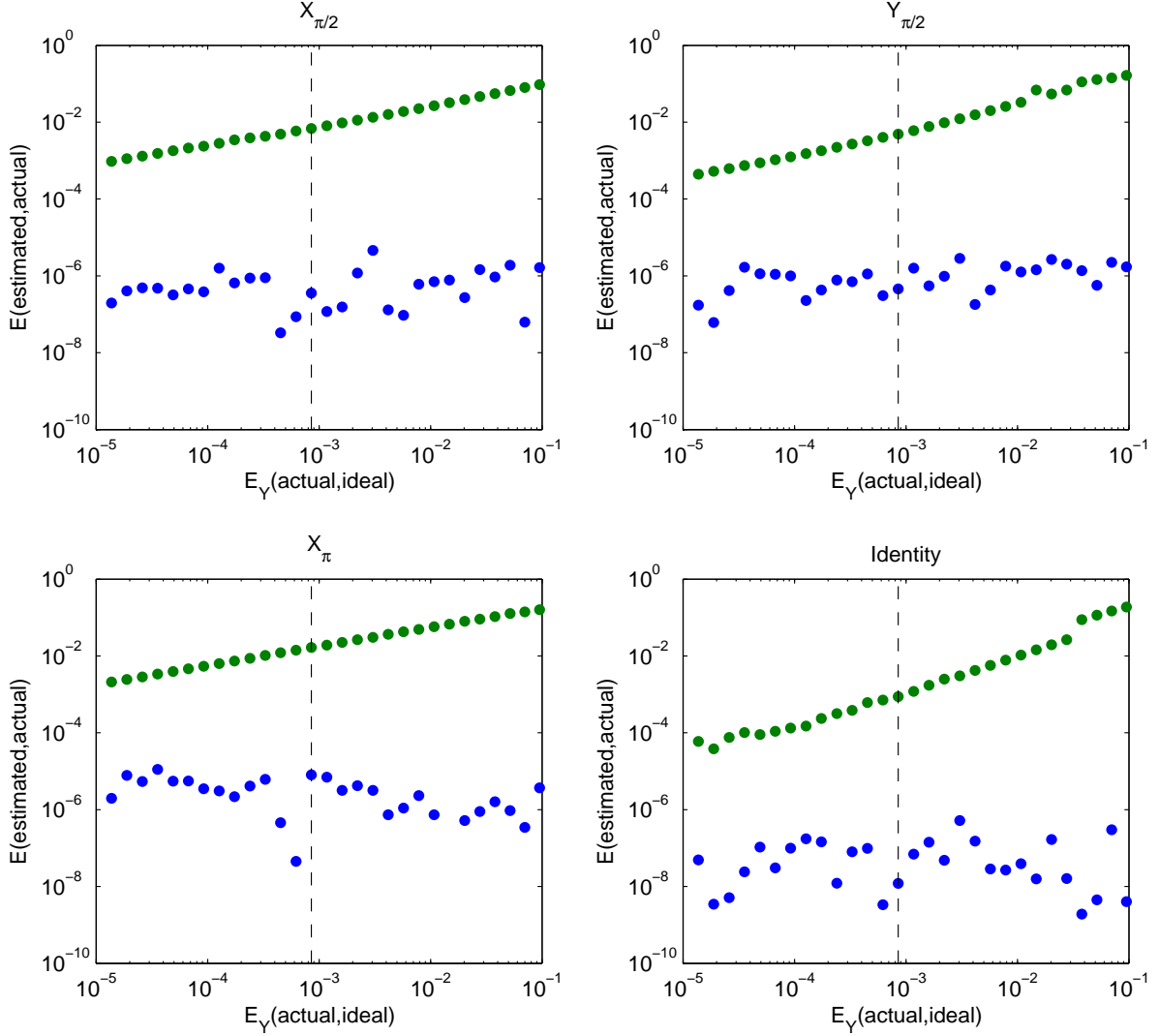}
\caption{Estimation error vs gate error for an over-rotation in $Y_{\pi/2}$. Blue dots are ML-GST and green dots are ML-QPT. The gate error is related to the over-rotation angle approximately as $E_Y = (\delta\theta_Y)^2/6$. The range of angles corresponding to the given range ($10^{-5} - 10^{-1}$) of gate error is $0.4\degree - 44.4\degree$. The vertical dashed line indicates the selected value of over rotation error ($4\degree$, $E = 8.5 \times 10^{-4}$) for which the PTMs are plotted on the following pages.} \label{est-vs-actual-error-theta-y}
\end{figure}

Fig.~\ref{est-vs-actual-error-theta-y} shows the estimation error of QPT and GST as a function of gate error for an over-rotation in the Y-gate, which is a type of coherent error. Gate error is defined as $\mathcal{E}_{gate} = 1-\bar{F}({\rm actual,ideal})$, where $\bar{F}({\rm actual, ideal})$ is the average fidelity of the actual gate relative to the ideal gate.\footnote{\label{average-fidelity-footnote}Average fidelity is defined as $\bar{F}(A,B) = \int d\rho \Tr\{\rho A^{-1}\circ B (\rho)\}$, where $A$ and $B$ are quantum operations, and the integral is over the uniform (Haar) measure on the space of density matrices. The average fidelity may be expressed in terms of Pauli transfer matrices $R_A, R_B$ as $\bar{F}(A,B) = (\Tr\{R_A^{-1}R_B\}+d)/[d(d+1)]$, where $d$ is the dimension of the Hilbert space. See Ref.~\cite{nielsen_simple_2002}.} Estimation error is defined similarly, as $\mathcal{E}_{est}=1-F({\rm estimated, actual})$. We are able to use fidelity as a metric because the estimates are constrained to be physical. (For unphysical gates, $0\leq \bar{F} \leq 1$ may not hold.) 

\begin{figure}
\includegraphics[width=16cm, trim = 1cm 0 0 0]{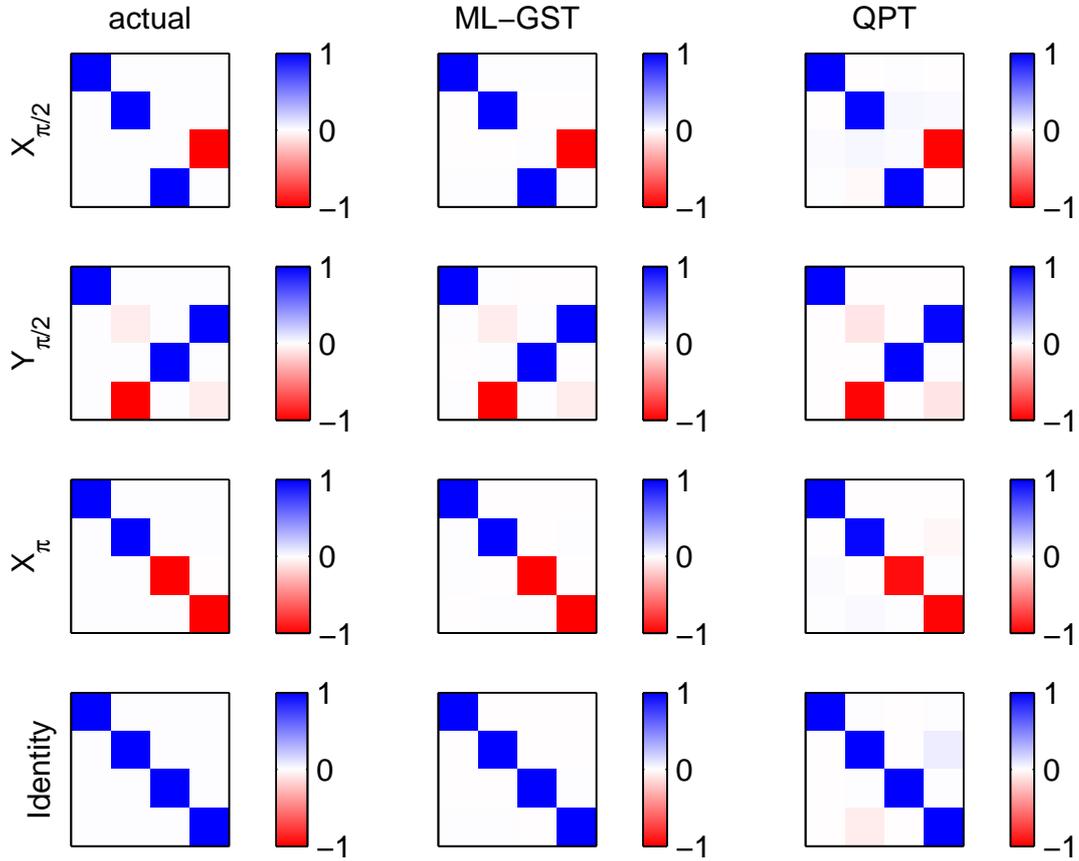}
\caption{Pauli transfer matrices for GST and QPT maximum likelihood estimates (2nd and 3rd columns). The actual gate set (1st column) contains an over-rotation error of $4\degree$ in the $Y_{\pi/2}$ gate, corresponding to a gate error (infidelity) of $E(\mathrm{actual},\mathrm{ideal}) = 8.5\times 10^{-4}$. GST correctly identifies the gate containing the error, while QPT attributes an over-rotation error to all gates.} \label{ptm-theta-y}
\end{figure}

\begin{figure}
\includegraphics[width=16cm, trim = 1cm 0 0 0]{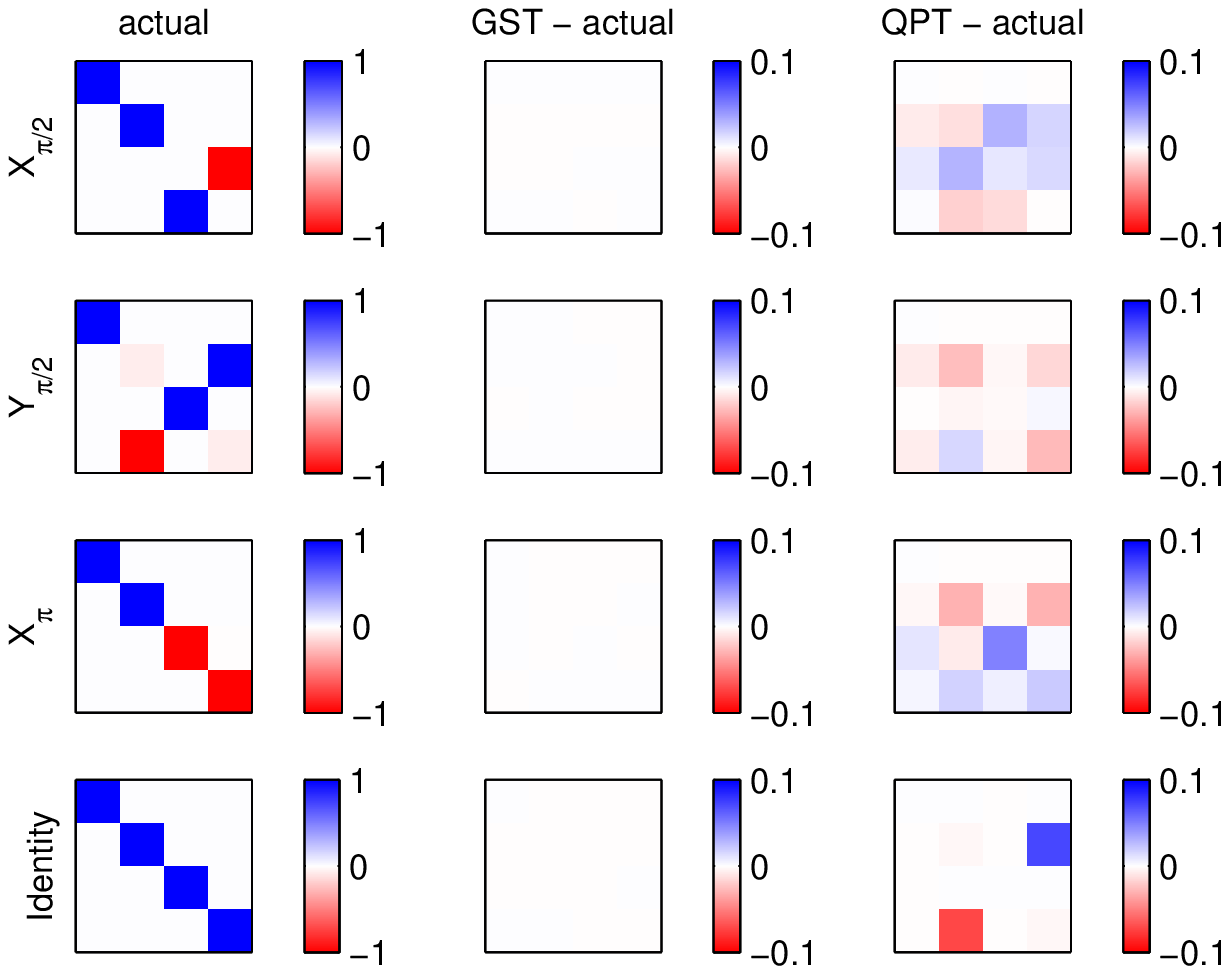}
\caption{Pauli transfer matrices for GST and QPT maximum likelihood estimates (2nd and 3rd columns) for the same data as in Fig.~\ref{ptm-theta-y}, with actual PTMs (column 1) subtracted off. The GST errors are about 1\% of QPT errors, and are not visible on this scale.} \label{ptm-diff-theta-y}
\end{figure}

We make the following observations: (1) QPT attributes error to all gates, even though only single gate ($Y_{\pi/2}$) is actually faulty. This is because the faulty gate is used in SPAM for all gates. (2) The GST estimation error is flat as a function of gate error, and represents the threshold for the optimizer.\footnote{\label{optimizer-threshold-footnote}We use a tolerance of $\epsilon = 10^{-12}$ on the (unweighted) least-squares objective function (see text), which corresponds to an estimation error of approximately $10^{-7}$. This can be derived by assuming an error in each entry of the PTM equal to the rms value of the residual in the objective function, $\sqrt{\epsilon/N_{exp}}$, where $N_{exp} \approx 10^2$ is the number of experiments. The estimation error is approximately equal to the error in the PTM elements, as can be verified using the trace formula in footnote~\ref{average-fidelity-footnote}, above.} (3) In the regime of gate error relevant to QEC thresholds ($\mathcal{E} = 10^{-3} - 10^{-2}$), (and given the numerical tolerances) GST is several orders of magnitude more accurate than QPT.

Fig.~\ref{ptm-theta-y} shows the PTM coefficients for the gate set at a representative gate error (infidelity) of $8.5 \times 10^{-4}$. Results are shown for the actual gates and for the GST and QPT estimates. Both GST and QPT are good at catching the error on the faulty ($Y_{\pi/2}$) gate, but QPT also attributes rotations to the other gates. The relative errors can also be seen in Fig.~\ref{ptm-diff-theta-y}, which shows the difference between the estimated and actual PTMs.

\subsubsection{Example 2: Depolarization (incoherent) error}

\begin{figure}
\includegraphics[width=16.4cm, trim = 2cm 0 0 0]{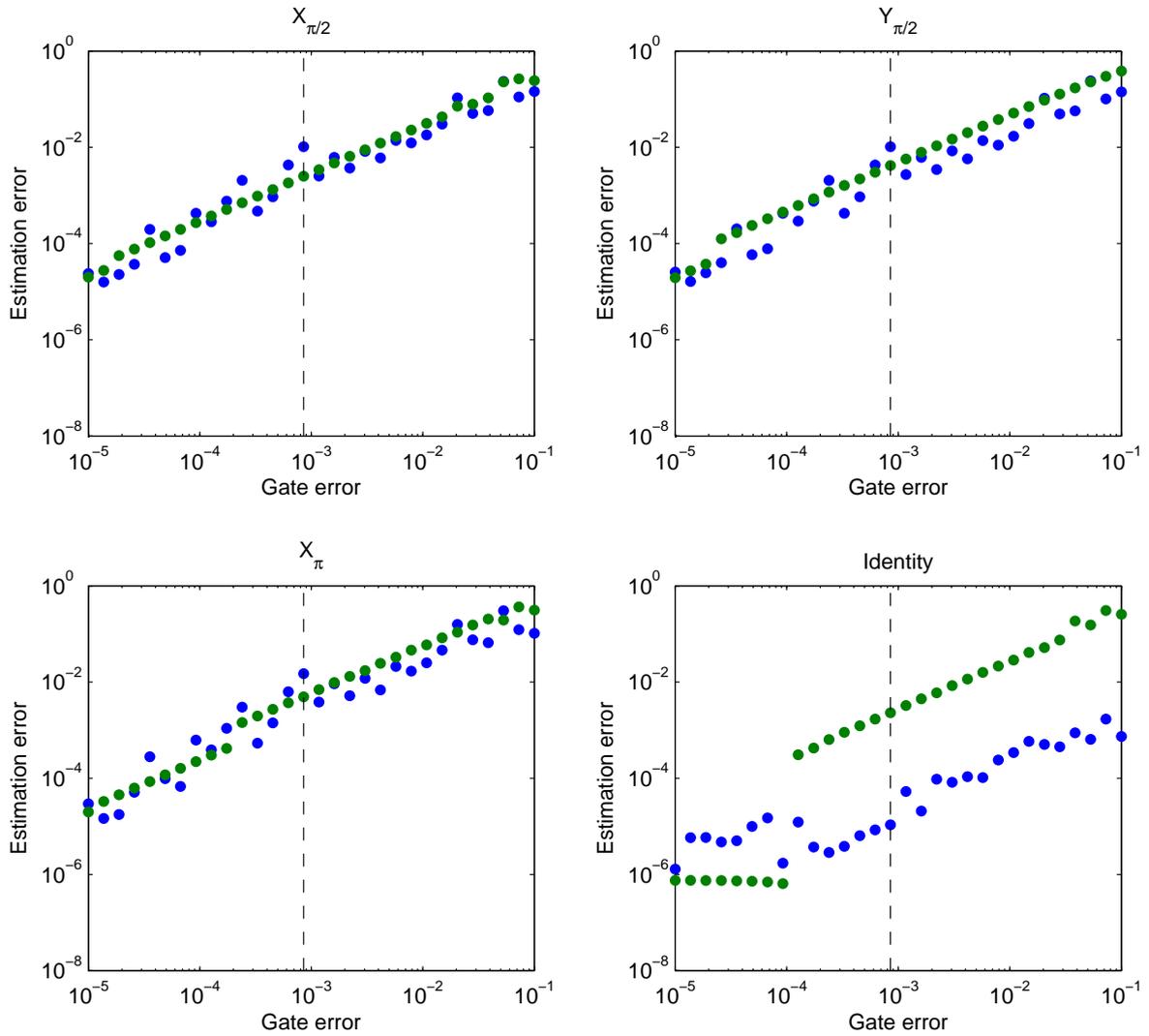}
\caption{Estimation error vs gate error for depolarizing noise on all gates. Blue dots are ML-GST and green dots are ML-QPT. Estimation error is given by the spectral norm of the difference between the estimated and actual PTMs and gate error is the infidelity between the actual and ideal PTMs. The gate error is proportional to the depolarizing parameter. The vertical dashed line indicates the selected value of gate error ($E = 8.5 \times 10^{-4}$) for which the PTMs are plotted on the following pages. This is the same error magnitude that was selected for the over-rotation example above. GST estimation error is about the same as QPT. See text for discussion.} \label{est-vs-actual-error-matnorm-depol}
\end{figure}

The depolarization map given by Eq.~(\ref{depol-gate}) is often used as a model of incoherent environmental noise. The depolarizing parameter $p$ is related to the gate error as $\mathcal{E} = 2p$.\footnote{This can be derived using the formula for average fidelity in footnote~\ref{average-fidelity-footnote}.} The estimation errors of QPT and GST as a function of gate error for depolarizing noise are plotted in Fig.~\ref{est-vs-actual-error-matnorm-depol}. As a measure of estimation error, we use the spectral norm (as implemented in Matlab with the function {\em norm.m}) of the difference between the estimated and actual gates. We use spectral norm rather than infidelity as in the previous example because the optimization routine was unable to satisfy the tolerance on the physicality constraints in the present case.\footnote{Technically, the diamond norm~\cite{KitaevBook} is a more correct metric for gate distance than the spectral norm. However, the spectral norm is easier to calculate. Since we are only interested in comparing the relative distance of QPT and GST estimates from the actual gate, and our simulated gates are {\em exactly} constrained to the single-qubit Hilbert space, the spectral norm is sufficient.} Although the constraint violation was small ($<10^{-4}$ deviation in each element the top row of the PTM), it was enough to invalidate fidelity as a metric.

As for the previous (over-rotation) example, we plot the PTMs for the estimated gates as well as the differences between estimated and actual PTMs for a representative gate error of $8.5 \times 10^{-4}$, the same as in the over-rotation example. At a given level of systematic noise, the estimation error of QPT is smaller for incoherent errors than coherent errors. 

According to Fig.~\ref{est-vs-actual-error-matnorm-depol}, the estimation errors of QPT and GST are about the same.\footnote{Based on the previous analysis of coherent errors, we expect the GST estimate not to vary as a function of the depolarizing parameter, since depolarizing noise is treated self-consistently in the same way. We expect this to be achievable using a more robust optimization routine than the one we use.} However, Fig.~\ref{ptm-diff-depol} shows that GST does a better job of estimating the non-zero coefficients of the PTM. These coefficients are $1-p$, where $p$ is the depolarizing parameter. Therefore GST gives a better estimate of $p$ than does QPT. This is one illustration of the danger in relying on a single parameter as a metric of gate or estimation quality.

\begin{figure}
\includegraphics[width=16cm, trim = 1cm 0 0 0]{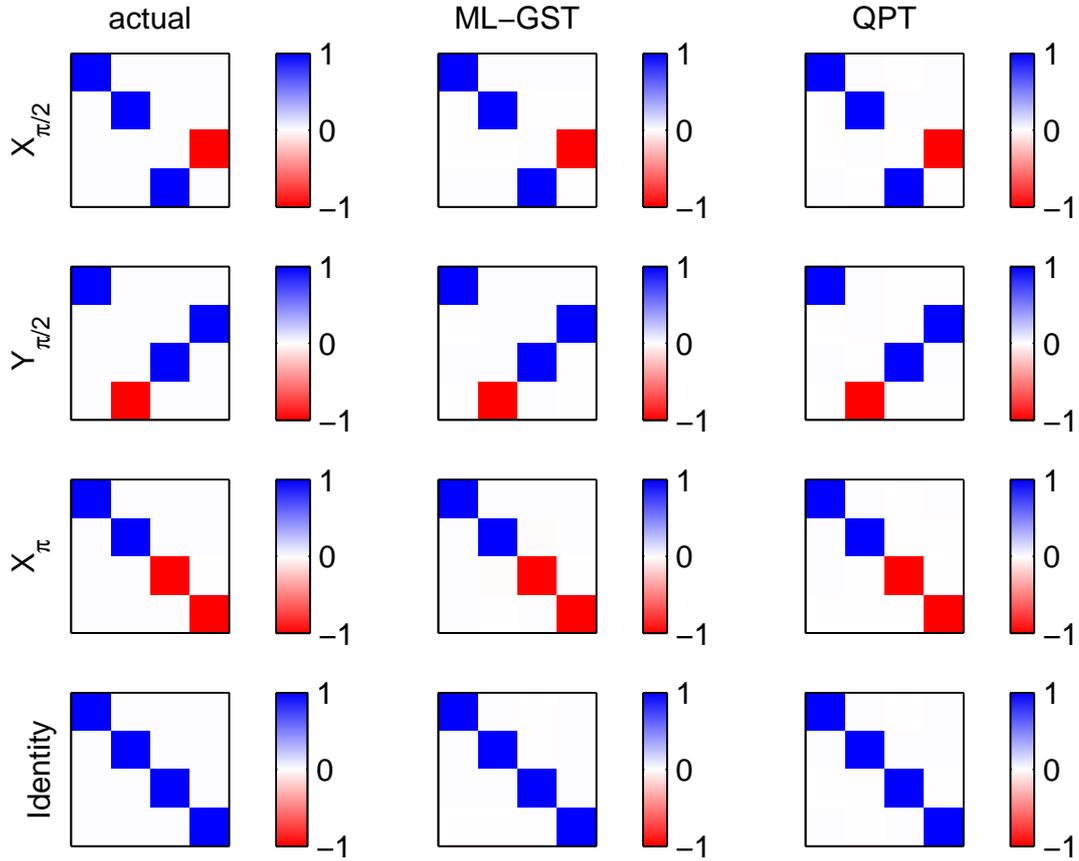}
\caption{Pauli transfer matrices for GST and QPT maximum likelihood estimates (2nd and 3rd columns). The actual gate set (1st column) contains a depolarizing gate error (infidelity) of $E(\mathrm{actual},\mathrm{ideal}) = 8.5\times 10^{-4}$, the same magnitude of gate error as in Fig.~\ref{ptm-theta-y}. QPT estimation error for depolarizing noise is smaller than that for over-rotation error and is not visible in this plot.} \label{ptm-depol}
\end{figure}

\begin{figure}
\includegraphics[width=16cm, trim = 1cm 0 0 0]{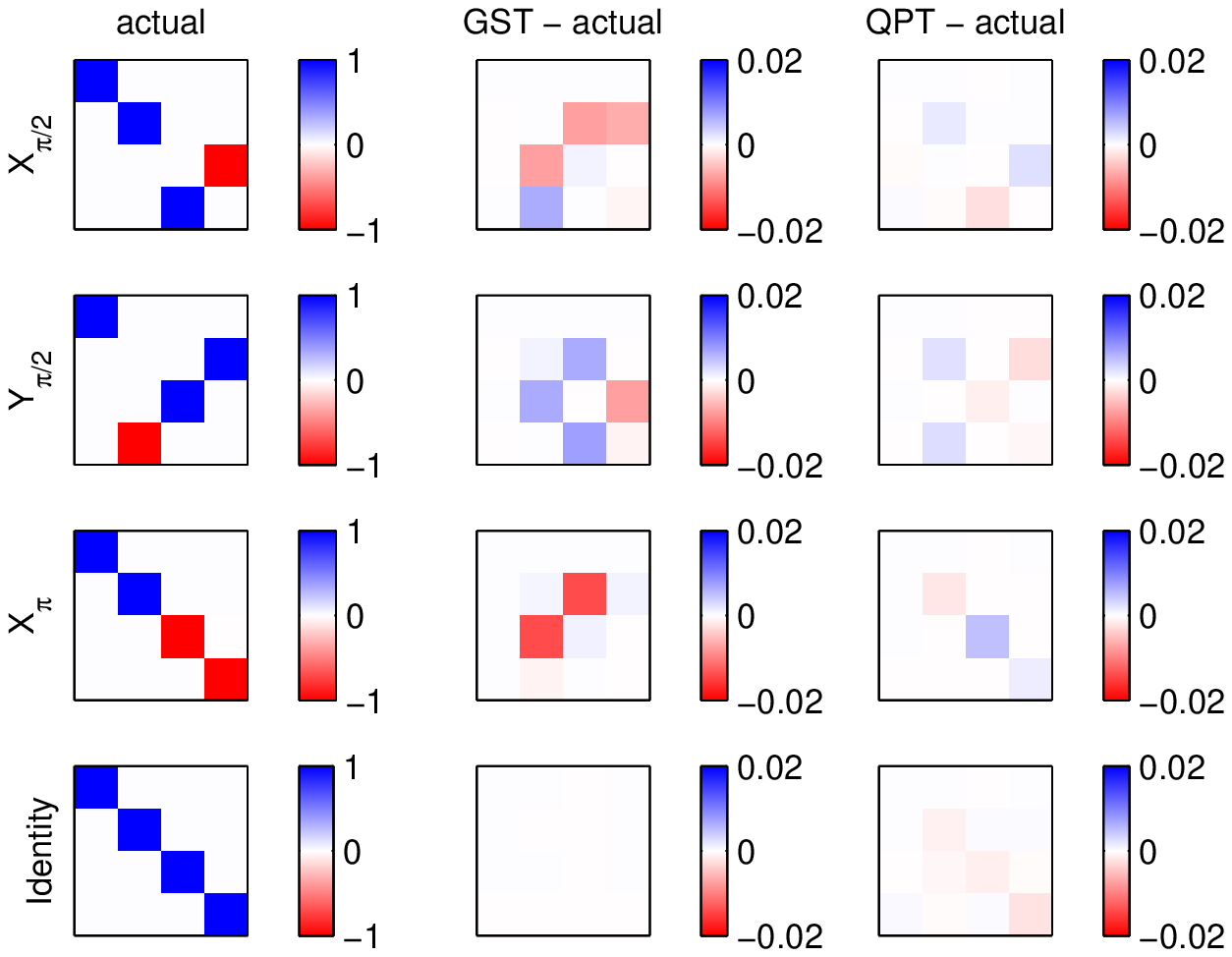}
\caption{Pauli transfer matrices for GST and QPT maximum likelihood estimates (2nd and 3rd columns) for the same data as in Fig.~\ref{ptm-depol}, with actual PTMs (column 1) subtracted off. The actual gate set (1st column) contains a depolarizing gate error (infidelity) of $E(\mathrm{actual},\mathrm{ideal}) = 8.5\times 10^{-4}$. The QPT estimation error is about an order of magnitude smaller overall for depolarizing noise than for over-rotation error, see Fig.~\ref{ptm-diff-theta-y}. The GST and QPT errors are comparable in magnitude, but QPT gives a worse estimate of the depolarizing parameter, which corresponds to the non-zero entries in the actual PTM.} \label{ptm-diff-depol}
\end{figure}

\subsubsection{Example 3: Intrinsic SPAM error}

As an example of intrinsic SPAM error, we assume a faulty initial state $\rho$, produced by applying depolarizing noise of strength $p$, Eq.~(\ref{depol-gate}), to an ideal initial state. $\lb E|\rho\rb = 0$ in the ideal case, and $\lb E|\rho\rb = 2 p$ is a measure of the state error. This is the same value as would be obtained for the gate error had the depolarizing noise been applied to the gate instead of the state. Although depolarizing noise commutes with all gates (the $i,j = 2\ldots d^2$ sub-matrix is proportional to the identity), the example considered here is not equivalent to that in the last section. This is because in the present case there is a single depolarizing operation acting in each experiment, whereas the number of depolarizing operations in the last section varied between 1 and 3, depending on which opperators appeared in $\lb E| F_i G_k F_j|\rho\rb$.

As could be expected from the results on depolarizing noise in the last section, the error in the QPT estimate grows linearly with the error on $\rho$. In fact, the QPT estimation error is almost exactly equal to the initial state error, showing that QPT attributes the noise to the gates rather than the state. This makes sense, since QPT assumes ideal initial states, and is confirmed by looking at the difference between the estimated and actual PTMs as in Fig.~\ref{ptm-diff-depol-rho}. In contrast, GST is insensitive to the initial state error, and the GST estimation error is at the optimizer threshold (see footnote~\ref{optimizer-threshold-footnote}).

\begin{figure}
\includegraphics[width=16.4cm, trim = 2cm 0 0 0]{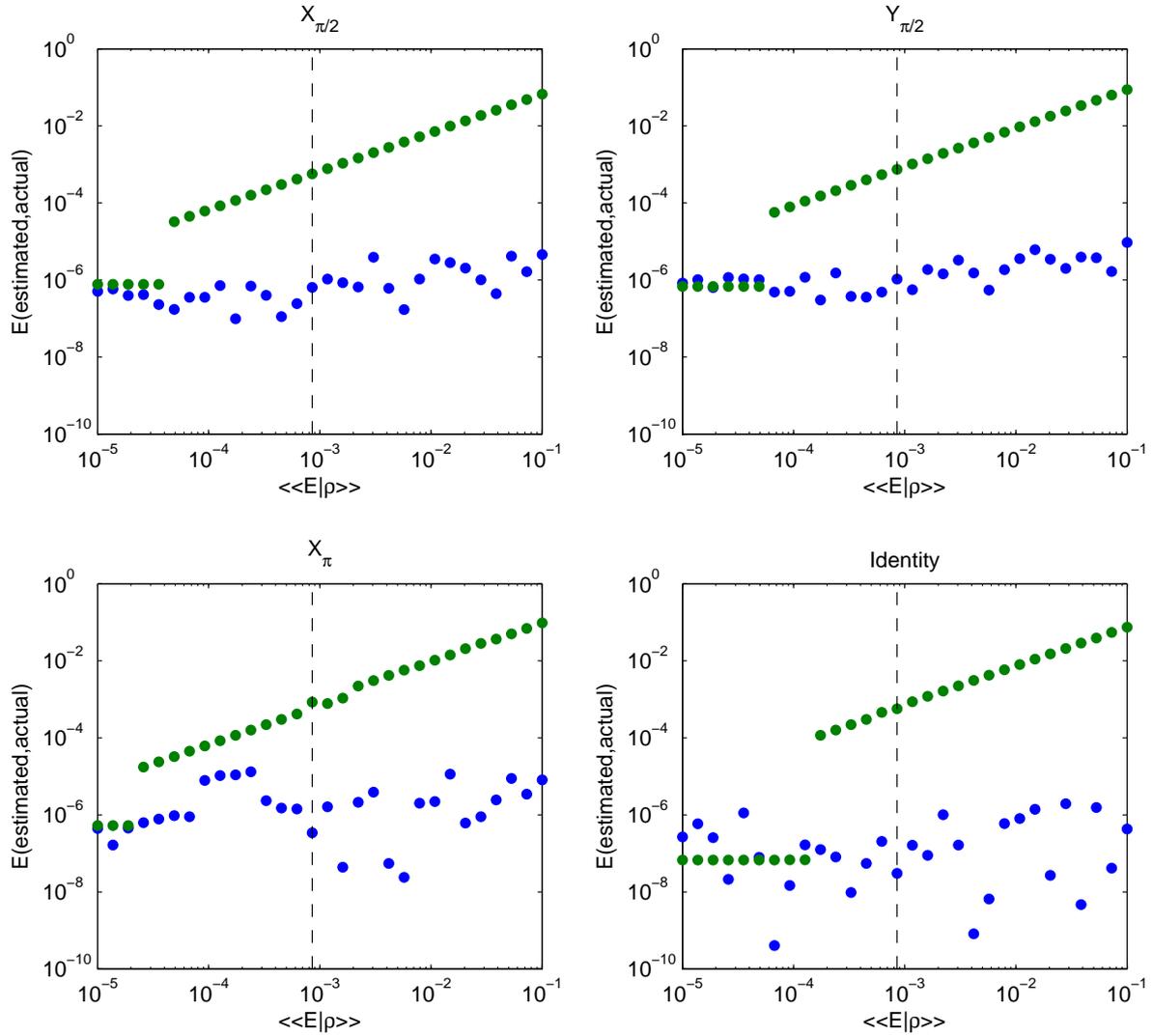}
\caption{Estimation error vs gate error for depolarizing noise on the initial state. Blue dots are ML-GST and green dots are ML-QPT. Estimation error is given by the infidelity as in Fig.~\ref{est-vs-actual-error-theta-y}. Initial state error is equal to $\lb E|\rho\rb$ in the present case, where $\lb E|$ and $|\rho\rb$ are orthogonal. The vertical dashed line indicates the selected value of state error ($E = 8.5 \times 10^{-4}$) for which the PTMs are plotted on the following pages. This is the same error magnitude that was selected for the over-rotation example above, and corresponds to what the infidelity would be if the depolarizing noise was attributed to the gate rather than the state.} \label{est-vs-actual-error-depol-rho}
\end{figure}

\begin{figure}
\includegraphics[width=16cm, trim = 1cm 0 0 0]{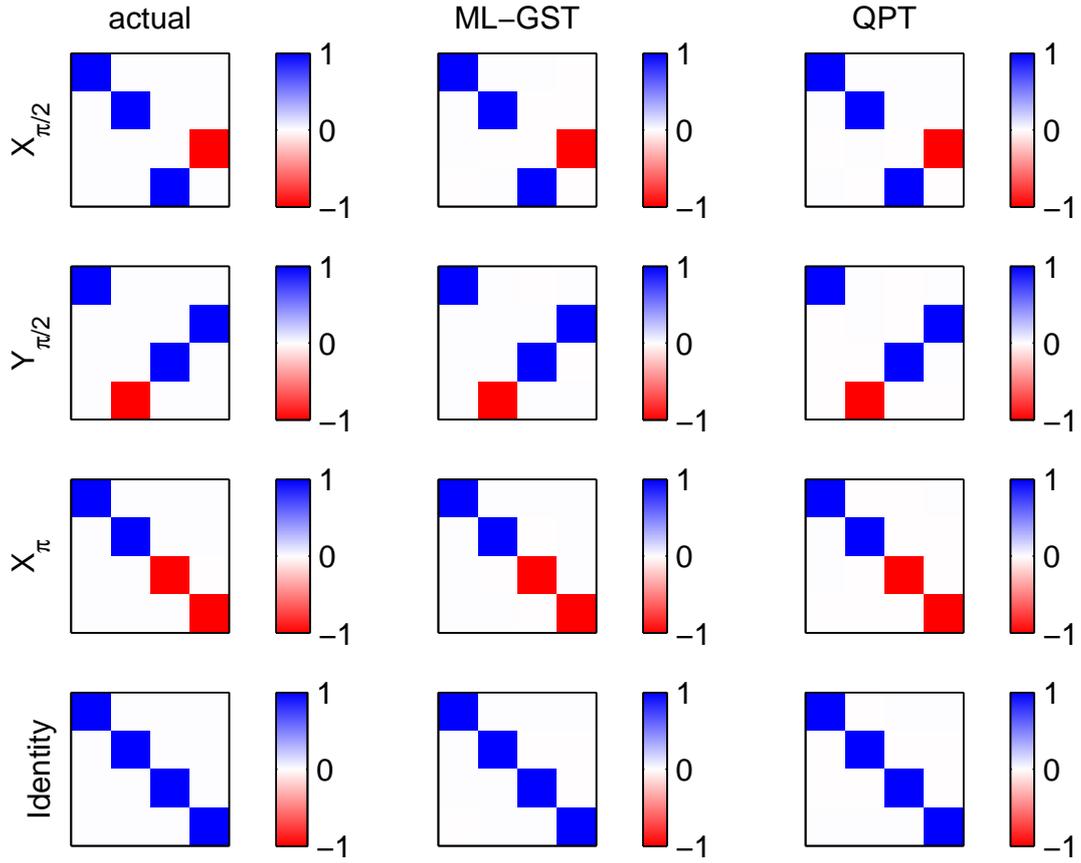}
\caption{Pauli transfer matrices for GST and QPT maximum likelihood estimates (2nd and 3rd columns). The actual gate set (1st column) contains no error. The error is a depolarizing noise of strength $\lb E|\rho\rb = 8.5\times 10^{-4}$ applied to the initial state.} \label{ptm-depol-rho}
\end{figure}

\begin{figure}
\includegraphics[width=16cm, trim = 1cm 0 0 0]{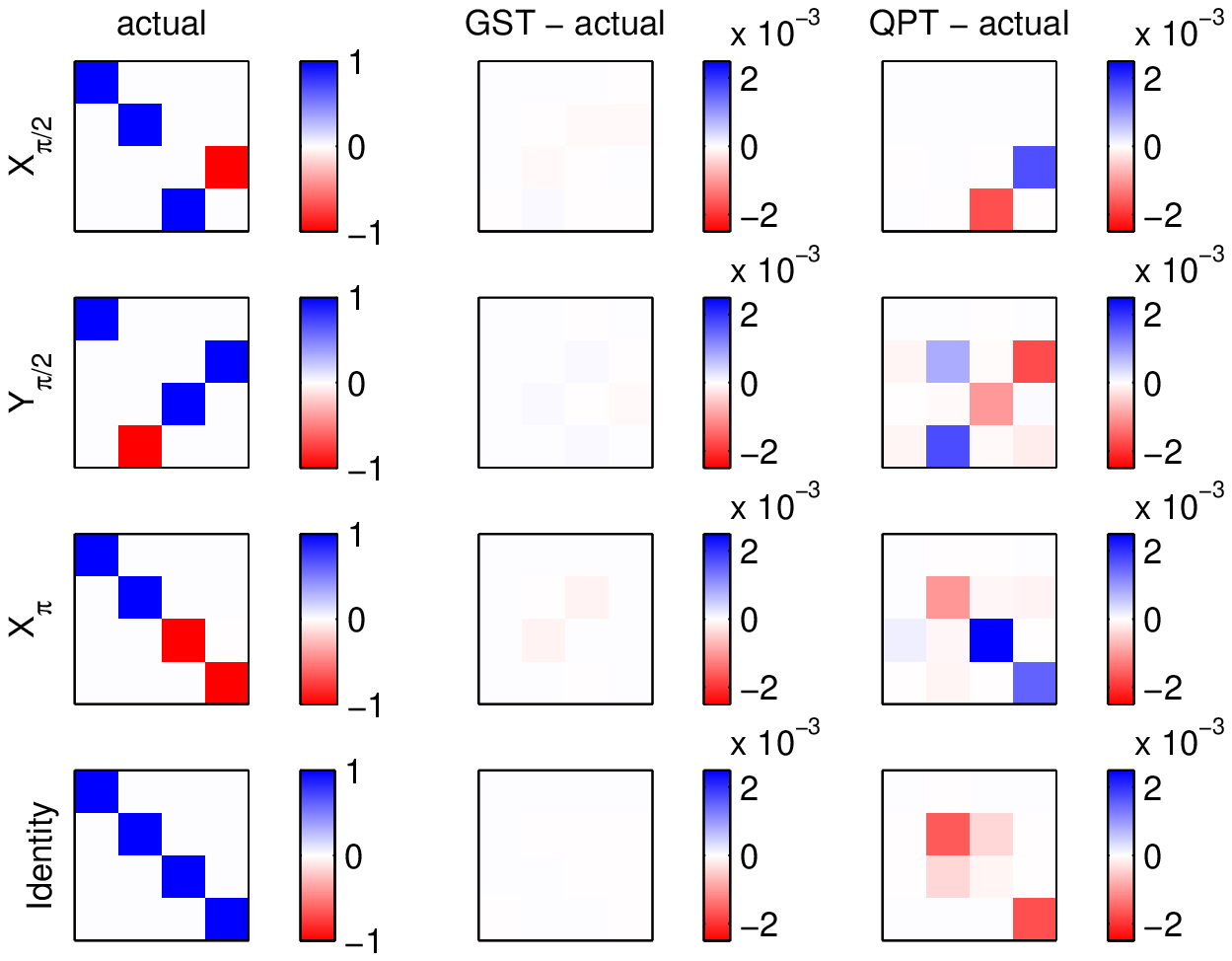}
\caption{Pauli transfer matrices for GST and QPT maximum likelihood estimates (2nd and 3rd columns) for the same data as in Fig.~\ref{ptm-depol-rho}, with actual PTMs (column 1) subtracted off. The error is a depolarizing noise of strength $\lb E|\rho\rb = 8.5\times 10^{-4}$ applied to the initial state. QPT incorrectly attributes a depolarization error to all gates.} \label{ptm-diff-depol-rho}
\end{figure}

\subsection{Sampling noise}

In this section we repeat the first example from the last section, over-rotations in the Y-gate, but this time in the presence of sampling noise and intrinsic SPAM error. This situation is closer to real life, and illustrates what happens in general. We take $N_{Samples} = 10000$, $\lb E|\rho\rb = 0.01$, both of which could reasonably occur in an experiment. This value of $N_{Samples}$ corresponds to a sampling error of about $0.01$ per PTM entry, so both errors are about the same order of magnitude. We expect that, as we vary the strength of the systematic error, QPT will be swamped by both sampling and intrinsic SPAM error until the systematic gate error rises above the level (0.01) of these errors. We expect GST, on the other hand, to be insensitive to intrinsic SPAM but to also be unable to detect the systematic error unless it is larger than the sampling error.

\begin{figure}
\includegraphics[width=16.4cm, trim = 2cm 0 0 0]{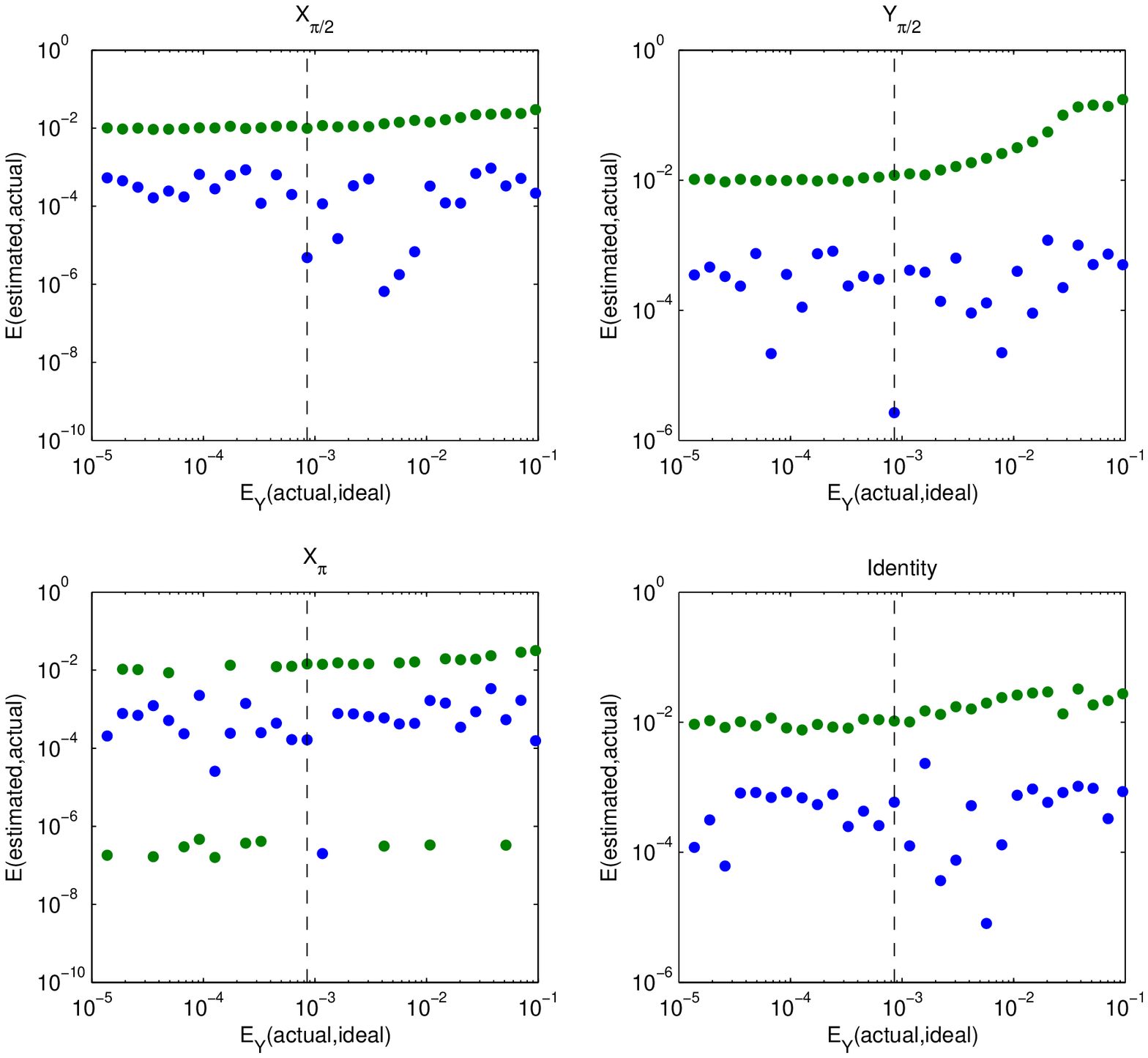}
\caption{Estimation error vs gate error for over-rotation in the Y-gate, including sampling noise and intrinsic SPAM errors with parameters $N_{Samples}=10000$, $\lb E|\rho\rb = 0.01$. Blue dots are ML-GST and green dots are ML-QPT. Estimation error is given by the infidelity as in Fig.~\ref{est-vs-actual-error-theta-y}. The vertical dashed line indicates the selected value of gate error ($E = 8.5 \times 10^{-4}$) for which the PTMs are plotted on the following pages.} \label{est-vs-actual-error-theta-y-samp}
\end{figure}

\begin{figure}
\includegraphics[width=16cm, trim = 1cm 0 0 0]{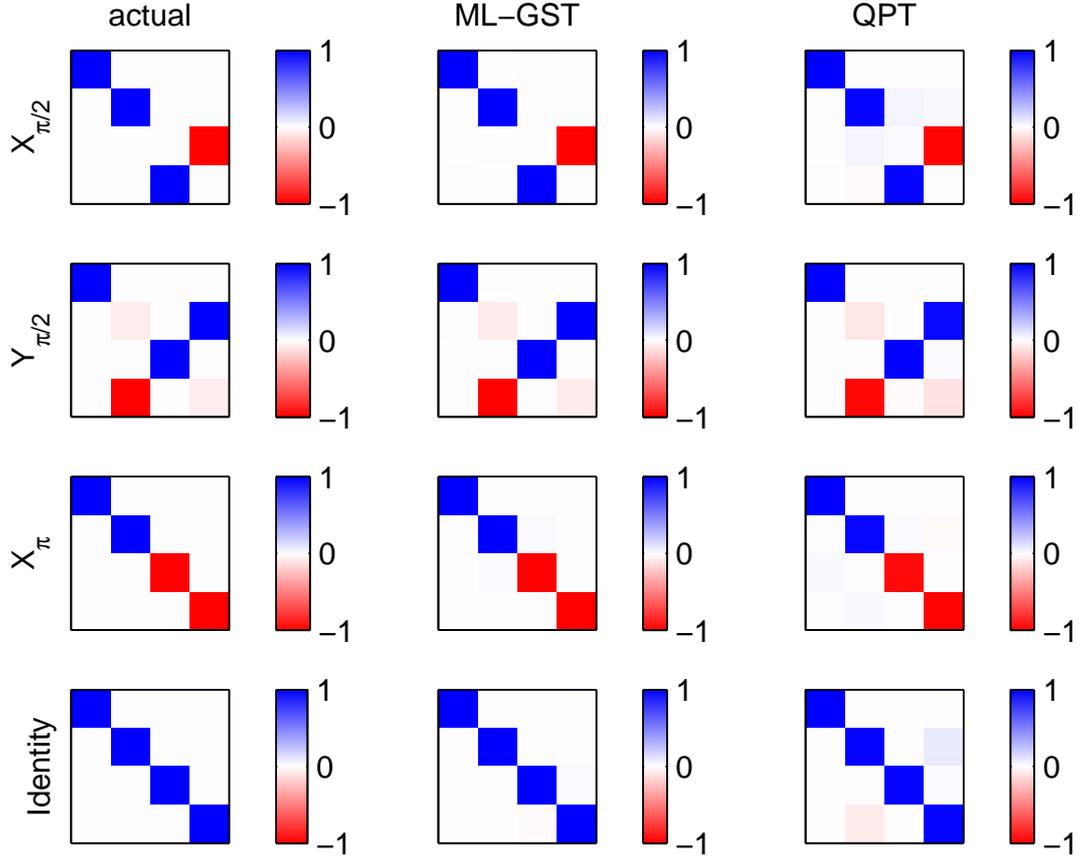}
\caption{Pauli transfer matrices for GST and QPT maximum likelihood estimates (2nd and 3rd columns) of the gate set in column 1. The actual gate set shown in column 1 contains an over-rotation error of $4\degree$ in the Y-gate, corresponding to a gate error of $8.5\times 10^{-4}$. Also input to the estimation were sampling noise and intrinsic SPAM errors with parameters $N_{Samples}=10000$, $\lb E|\rho\rb = 0.01$.} \label{ptm-depol-theta-y-samp}
\end{figure}

\begin{figure}
\includegraphics[width=16cm, trim = 1cm 0 0 0]{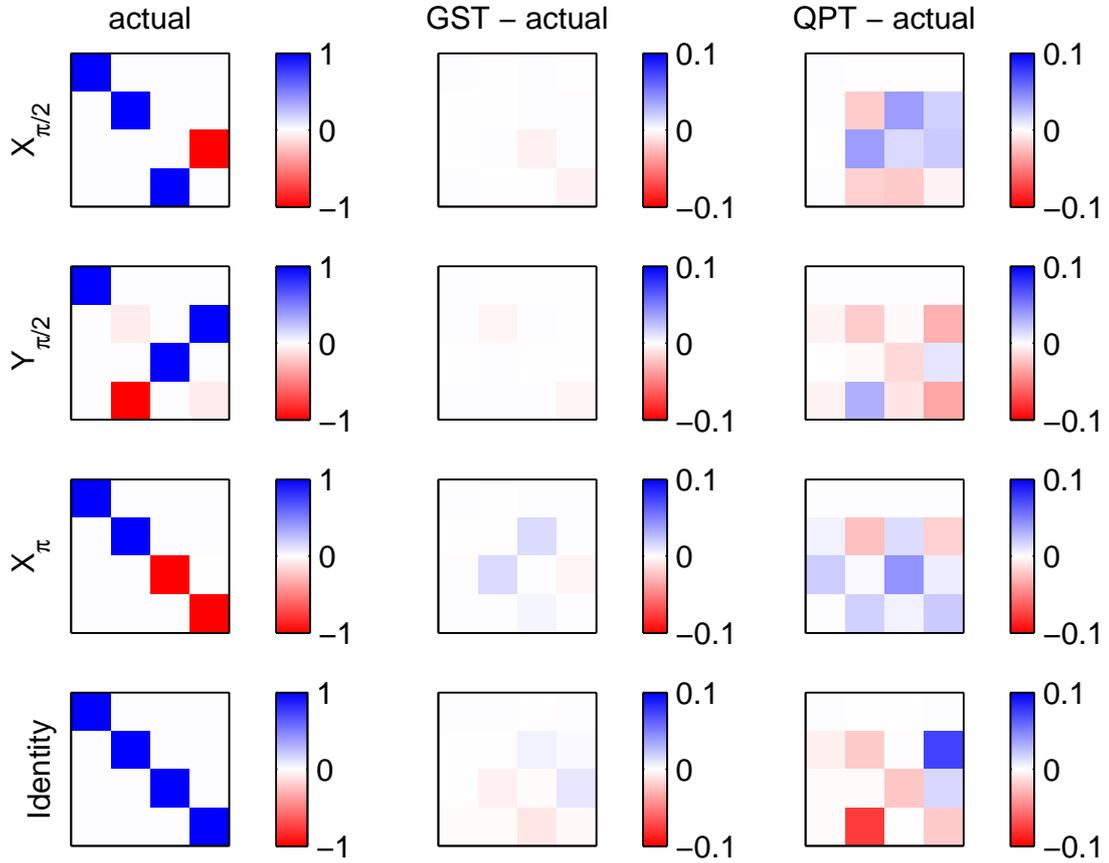}
\caption{Pauli transfer matrices for GST and QPT maximum likelihood estimates (2nd and 3rd columns) for the same data as in Fig.~\ref{ptm-depol-theta-y-samp}, with actual PTMs (column 1) subtracted off. The actual gate set (1st column) contains an over-rotation error of $4\degree$ in the Y-gate, corresponding to a gate error of $8.5\times 10^{-4}$. Also input to the estimation were sampling noise and intrinsic SPAM errors with parameters $N_{Samples}=10000$, $\lb E|\rho\rb = 0.01$.} \label{ptm-diff-depol-theta-y-samp}
\end{figure}

The results turn out to be more subtle. Depending on the type of error, it may be detectable at a lower gate error than the sampling error. This is the case for over-rotations. Fig.~\ref{est-vs-actual-error-theta-y-samp} shows the estimation error vs gate error for our example. As expected, the estimation error is flat until the gate error exceeds 0.01, both for QPT and GST. QPT is limited by intrinsic SPAM while GST is not. Above 0.01, the estimation error behaves as in Fig.~\ref{est-vs-actual-error-theta-y} -- the QPT values increase while GST remains flat. However, the PTMs give a fuller picture. Figs.~\ref{ptm-depol-theta-y-samp} and~\ref{ptm-diff-depol-theta-y-samp} show the PTMs for a gate error near $10^{-3}$, well below the crossover point. GST is still able to find the over-rotation error. This is because, for a given gate error, the magnitude of non-zero PTM entries due to the over-rotation error is larger than the average magnitude of PTM entries for sampling noise.

The results in this section were generated by numerical optimization in Matlab, as in Sec.~\ref{examples-subsection}. Everything stated in that section regarding the optimization carries over to here, except that the objective function used was weighted least squares, and the objective function tolerance was set to options.TolFun = $10^{-6}$. The reason is that in the presence of sampling noise, the weighted-least squares objective function is equal to the chi-squared, which is of order $N_{exp}$. Without sampling noise, the unweighted least squares objective function is near zero, hence the smaller function tolerance in that case.

\chapter{Summary and outlook}

We have seen that gate set tomography is a robust and powerful tool for the full characterization of quantum gates. In the presence of SPAM errors, GST is accurate (within the limits of sampling error) while QPT typically overestimates the gate error. This discrepancy can be several orders of magnitude in the regime of gate error relevant to quantum error correction. GST is also capable of providing {\em qualitative} information about systematic gate errors -- via the estimated Pauli transfer matrices -- that is not accessible by QPT in the presence of SPAM error. Nevertheless, several topics remain the subject of current research. These include reduction of sampling error, treatment of non-Markovian noise, detection of leakage and extra dimensions, and tractable extension to multiple qubits.

Of these, the extension to multiple qubits seems the most clear-cut. For multiple qubits, the formalism of GST is the same but there is a formidable processing challenge because of the large amount of experimental data necessary (over 4,000 experiments for 2-qubit GST, as compared to $<100$ for 1-qubit). The nonlinear optimization routines we used to illustrate GST in Ch.~\ref{implementation-chapter} would likely take an unreasonable amount of time even for 2 qubits, assuming they run at all. The SDP method discussed in Ch.~\ref{derivation-chapter} has been reported to run slowly for 2 qubits as well~\cite{IBMpcomm}. One solution to this problem has been proposed by IBM, which is to use a type of semidefinite program used for compressed-sensing problems, called a first-order conic solver. Preliminary reports indicate that this runs much faster than standard SDP~\cite{IBMpcomm}.

Robin Blume-Kohout has proposed the inclusion of gate repetitions in the GST gate set in order to reduce sampling error and detect non-Markovian noise~\cite{RBK-pcomm}. Some initial evidence of the usefulness of this approach was presented in Ref.~\cite{RBK2013}. The detection of extra dimensions -- i.e. due to the system the leaving the qubit Hilbert space -- is also an active topic of research~\cite{RBK-pcomm,stark_compressibility_2014}.

One topic we have not discussed in detail but that deserves fuller attention is estimating the error in the GST estimates. This is important in order to determine the resource requirements (number of experiments, gate repetitions, etc.) for obtaining high-quality GST estimates within the tolerances required by QEC. It is a pressing question whether the required accuracy can be tractably obtained for two-qubit gates. E.g., for a CNOT gate below the surface code threshold -- probability of gate error $ \approx 10^{-3}$ -- we would like the error in the GST estimate of this quantity to be below roughly $10^{-4}$. 

Part of the error in the maximum-likelihood estimate comes from sampling noise, part from any approximations used in the objective function, and part from the optimization routine (which caused the point-to-point variability in the plots in Ch.~\ref{implementation-chapter} of estimation vs gate error in the absence of sampling error). Neglecting errors due to the optimization itself, it is important to understand how statistical sampling errors propagate through the MLE procedure. Standard errors for MLE estimates (see, e.g. Ref.~\cite{statistics-book}, Ch.~14) are valid when these estimates can be shown to be asymptotically (limit of large number of experiments) normally distributed. In this case, an asymptotically valid estimator for the error can be written in terms of the covariance matrix of the estimated parameter vector $\hat{\vec{t}}$, which can be calculated from the MLE estimate and the objective function. However, the MLE estimates for GST may not be asymptotically normally distributed due to the constraint that $\vec{t}$ must describe a physical map. In fact, for very small errors - the regime we are interested in - the gates will be very close to ideal unitaries, which lie on the boundary of allowable maps. Therefore asymptotic normality cannot be assumed. Hence it is unclear how accurate this approach would be for our problem.

A commonly used~\cite{Chow2012SI} approach to estimating the error in the ML estimate is to resample from experimental data, known as bootstrapping. In Ch.~\ref{implementation-chapter}, we were able to plot estimation error vs gate error because we knew what the actual gates were. However, the goal of GST is to estimate an unknown set of gates. Operationally, the variance in the estimate can be found by repeating the same experiment many times, each time generating new data and a new best fit. Since the amount of work to do this can be impractical (and if it isn't one would prefer to include this additional data in a single, larger sample to produce a tighter estimate), resampling with replacement from the same data (say a set of single-shot measurements of size $N_{Samples}$) seems like a good alternative. Unfortunately, it is well known~\cite{blume-kohout_robust_2012} that bootstrapping is unreliable for biased estimators. As discussed in the previous paragraph, the MLE estimate is biased due to physicality constraints. Therefore, without a rigorous theory of error estimation for quantum tomography in the presence of sampling noise, it is impossible to evaluate the validity of bootstrapping for this problem.

A practical solution to both of these problems (validity of standard errors for MLE, validity of bootstrapping) may be possible via a Monte Carlo approach. One can numerically generate many sample data sets with a specified error model and a given level of  sampling noise, as we have done in Ch.~\ref{implementation-chapter}. One can then perform GST on each sample data set and study the distribution of the resulting estimates. It is possible that general features, such as the amount of bias in the GST estimates for a given model of systematic errors, may be obtained in this way. This can then be used to determine under what conditions techniques such as bootstrapping are justified.

In conclusion, we have presented an overview of gate set tomography for a single qubit. It is hoped that this will be useful to practitioners aiming to implement full characterization of single as well as multi-qubit gates.

\chapter{Acknowledgements}

I would like to thank Robin Blume-Kohout, Jay Gambetta, Easwar Magesan, Andrew Skinner, Yudan Guo, Ben Palmer, Tim Sweeney, Ramesh Bhandari, and Michael Mandelberg for helpful discussions and input.

\bibliography{mybib}{}
\bibliographystyle{apsrev}

\end{document}